\documentclass[%
reprint,
superscriptaddress,
amsmath,amssymb,
aps,
prb,
]{revtex4-2}

\usepackage{graphicx}
\usepackage{dcolumn}
\usepackage{bm}
\usepackage{amsthm}
\usepackage[colorlinks,bookmarks=true,citecolor=blue,linkcolor=red,urlcolor=blue]{hyperref}
\usepackage{bbm}
\usepackage[subrefformat = parens, caption = false, labelformat=parens]{subfig}
\usepackage{floatrow}
\usepackage{verbatim}
\usepackage{overpic}

\theoremstyle{definition}
\newtheorem{protocol}{Protocol}

\makeatletter
\renewcommand\p@subfigure{\thefigure}
\makeatother

\usepackage[acronym]{glossaries}
\makenoidxglossaries

\newacronym{mzms}{MZMs}{Majorana zero modes}
\newacronym{dos}{DOS}{density of states}
\newacronym{gf}{GF}{Green's function}
\newacronym{wrt}{w.r.t.}{with respect to}

\glsdisablehyper

\newcommand{\eq}[1]{Eq.~(\ref{#1})}

\begin{document}

\preprint{APS/123-QED}

\title{Three-Majorana Cotunneling Interferometer for Non-Abelian Braiding and Topological Quantum Gate Implementation}

\author{Zhen Chen}
\affiliation{International Center for Quantum Materials, School of Physics, Peking University, Beijing 100871, China}

\author{Yijia Wu}
\thanks{Corresponding author: yijiawu@fudan.edu.cn}
\affiliation{Interdisciplinary Center for Theoretical Physics and Information Sciences, Fudan University, Shanghai 200433, China}
\affiliation{State Key Laboratory of Surface Physics and Institute for Nanoelectronic Devices and Quantum Computing, Fudan University, Shanghai 200433, China}
\affiliation{Hefei National Laboratory, Hefei 230088, China}

\author{X. C. Xie}
\affiliation{International Center for Quantum Materials, School of Physics, Peking University, Beijing 100871, China}
\affiliation{Interdisciplinary Center for Theoretical Physics and Information Sciences, Fudan University, Shanghai 200433, China}
\affiliation{Hefei National Laboratory, Hefei 230088, China}

\date{\today}

\begin{abstract}
We propose a novel scheme for performing Majorana zero mode (MZM) braiding utilizing cotunneling processes in a three-MZM system incorporating reference arms. 
This approach relies on the interference between cotunneling paths through the MZMs and reference arms, establishing an effective, tunable coupling between the MZMs. 
The strength and sign of this coupling can be manipulated via the reference arms and applied magnetic flux. 
Notably, the introduction of a half quantum flux reverses the coupling sign, enabling an echo-like protocol to eliminate dynamic phases during braiding. 
Our setup, requiring only three MZMs, represents a minimal platform for demonstrating non-Abelian braiding statistics. 
We demonstrate that this system facilitates the implementation of Clifford gates via braiding and, significantly, permits the realization of non-Clifford gates, such as the $T$ gate, by geometric phase, thereby offering a potential pathway towards universal topological quantum computation.
\end{abstract}

\maketitle


\section{Introduction} \label{sec:introduction}
Majorana zero modes (MZMs), exotic quasiparticles emerging as zero modes of topological superconducting systems whose anti-particles are themselves, have garnered significant attention due to their potential application in fault-tolerant topological quantum computation \cite{majorana1937Teoria, Ivanov2001Non, kitaev2001Unpaired, kitaev2003Fault, nayak2008Non, trebst2008Short, lutchyn2010Majorana,oreg2010Helical, alicea2011non, alicea2012new, zhang2019Next, beenakker2020Search, yazdani2023Hunting}. 
Two spatially separated MZMs can combine to form a nonlocal Dirac fermion, enabling the encoding of quantum information in a topologically protected manner. 
This nonlocal encoding, distributed across physically distant MZMs, inherently protects the quantum information from local perturbations and decoherence mechanisms. Furthermore, the non-Abelian exchange statistics of MZMs allow for the implementation of quantum gates through braiding operations—the systematic exchange of MZMs' positions in real space. These topologically protected operations, arising from the system's underlying geometry rather than precise Hamiltonian control, offer a pathway to fault-tolerant quantum computation that is intrinsically robust against local environmental noise.

Various platforms and protocols for MZM braiding have been proposed \cite{fu2008Superconducting,leijnse2012Parity, nadjperge2014Observation, pientka2017Topological, fornieri2019Evidence, ren2019Topological, livanas2019Alternative, prada2020Andreev,chen2020Atomic, thomson2022Gate, dvir2023Realization, aghaee2023Inas, marra2024Majorana, seoanesouto2024Subgap, torresluna2024Flux, tenhaaf2024Two, kezilebieke2020Topological}, and transport signatures of MZMs have been extensively explored \cite{law2009Majorana, flensberg2010Tunneling, liu2012Zero, kells2012Near, albrecht2016Exponential,rosdahl2018Andreev, mnard2020Conductance, valentini2022Majorana,pan2022Demand, hess2023Trivial}.
One attractive approach is to use one-dimensional topological superconductors, 
such as semiconductor nanowires proximitized by s-wave superconductors \cite{lutchyn2010Majorana, oreg2010Helical, mourik2012Signatures, deng2016Majorana, lutchyn2018Majorana, prada2020Andreev,vanloo2023Electrostatic, ten2025observation}. 
For such one-dimensional systems, many braiding schemes rely on Y-junction or tri-junction geometries, 
which typically necessitate the spatial moving of at least four MZMs to perform a braid \cite{alicea2011non, harper2019Majorana, torres2024Design,karzig2017scalable}. 
Alternative implementations include two-dimensional platforms, such as MZMs bound to vortices in iron-based superconductors \cite{wang2018Evidence, zhang2018Observation, kong2019Half, zhu2020Nearly, hu2024Dislocation}, 
which operate on the same topological mechanisms as one-dimensional chains but offer different experimental platforms.
Usually, these systems' conventional braiding schemes rely on physically moving MZMs in real space, which introduces 
significant experimental complexities in fabrication, control precision, and decoherence management. 

To overcome these weaknesses, other braiding methods like Coulomb-assisted braiding \cite{vanHeck2012Coulomb} or measurement-based protocols \cite{Bonderson2008Measurement, Bonderson2013Measurement, vijay2016Teleportation, alan2020optimizing} have been proposed. 
In this work, we introduce and analyze a new approach to MZM braiding utilizing electron cotunneling processes within three spatially separated MZMs ($\gamma_0, \gamma_1, \gamma_2$).
This approach offers experimental convenience by allowing the same apparatus to be used for both qubit/parity measurements \cite{fu2010Electron,whiticar2020coherent} and braiding operations. This is particularly timely given recent experimental advances in interferometric single-shot parity measurements in InAs-Al hybrid devices \cite{aghaee2025Interferometric}.
In our approach, each MZM is connected to a lead, and cotunneling dominates the transport through a pair of MZMs due to the Coulomb blockade effect. 
Electron current and noise studies of such a system have been performed in Ref.~\cite{fu2010Electron,albrecht2016Exponential,valentini2022Majorana, whiticar2020coherent, manousakis2020Weak}. 
We demonstrate that when this 3-MZM cotunneling system is incorporated into an interferometry setup with reference arms connecting the leads, an effective coupling between pairs of MZMs arises due to cotunneling interference effects. 
This effective coupling, shown as the Lamb term \cite{Breuer2002theory,Wiseman2009Quantum} $H_{LS} \sim i \Delta_{\alpha \alpha'} \gamma_\alpha \gamma_{\alpha'}$ in the Lindblad master equation describing the system, 
can be precisely controlled in magnitude and sign by manipulating the transmission through the reference arms or, 
significantly, by tuning the magnetic flux $\Phi$ threading the interference loops.
Such controllable coupling provides us with the flexibility to implement non-Abelian braiding statistics.
As for the braiding process, we find that a crucial challenge in braiding schemes involving direct MZM hybridization is that both the desired topological (geometric) phase and the unwanted dynamic phase are simultaneously accumulated. 
To solve this problem, we leverage the flux tunability of our cotunneling-mediated coupling to implement a spin echo-like technique. 
By applying a half-quantum flux ($\Phi_0/2 = h/2e$) midway through a braiding cycle, 
the sign of the effective MZM coupling is reversed ($t_{R,\alpha \alpha'} \rightarrow -t_{R,\alpha \alpha'}$), 
allowing the dynamic phase acquired in the first half of the operation to be cancelled by the dynamic phase accumulated in the second half. 
In this way, the braiding outcome arises entirely from the non-Abelian geometric transformation associated with the braiding properties of MZMs.

Furthermore, achieving universal quantum computation requires supplementing the Clifford gates generated by braiding with at least one non-Clifford gate, 
such as the $T$ gate ($\pi/8$ phase gate). 
Such a $T$ gate is usually generated by dynamic phase since pure physical space exchange braiding method of MZMs is prohibited to generate non-Clifford gate\cite{sarma2015majorana,beenakker2020Search}. 
This reliance on dynamic phase becomes the Achilles' heel of the MZM-based system, leaving it exposed to decoherence and unprotected by the topological mechanisms that shield the Clifford operations\cite{hyart2013Flux,sarma2015majorana,karzig2017scalable}.
To improve this, we show that our 3-MZM cotunneling platform, by enabling precise control over the MZM couplings 
($ \Delta_{01}, \Delta_{02}, \Delta_{12} $) via the reference arms and flux, allows for the generation of the $T$ gate purely through 
geometric phase manipulation. By guiding the system along a specific closed path in the coupling parameter space using an echo sequence 
to cancel the dynamic phase, a desired geometric phase corresponding to the $T$ gate can be accumulated in a controlled manner.

We develop the theoretical framework for this cotunneling-assisted braiding scheme using second-order perturbation theory to derive the effective MZM Hamiltonian. 
We further analyze the robustness of the proposed protocols by employing a quantum master equation approach and present numerical simulations that demonstrate the feasibility of performing high-fidelity Clifford gates and the non-Clifford $T$ gate using the braiding protocols.

This paper is organized as follows: Section~\ref{sec:hamiltonian} details the model Hamiltonian and the derivation of the effective cotunneling-induced MZM couplings. 
Section~\ref{sec:braiding} outlines the braiding protocol using the echo technique for dynamic phase cancellation and presents simulation results based on the master equation. 
Section~\ref{sec:tgate} describes the procedure for generating the $T$ gate via geometric phase control with simulation results. 
Finally, Section~\ref{sec:conclusion} provides a summary and concluding remarks.

\section{Model and Hamiltonian} \label{sec:hamiltonian}
\subsection{Model Hamiltonian}
\begin{figure}[htp!]
\centering
\subfloat[]{
\begin{overpic}[width=0.8\linewidth]{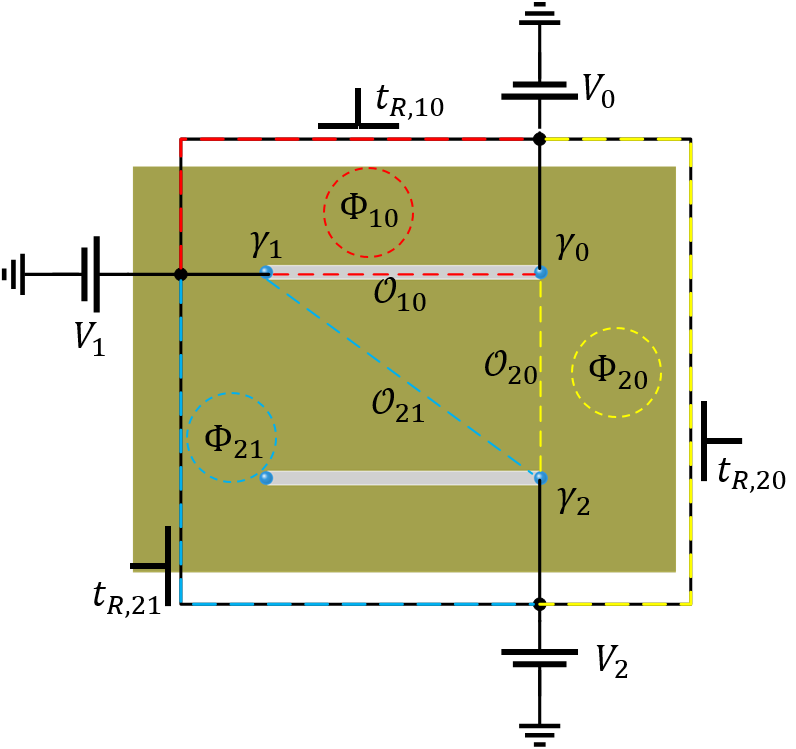}
\put(0,90){\textbf{(a)}}
\end{overpic}
\label{fig:setup_2d}
}
\\
\subfloat[]{
\begin{overpic}[width=0.85\linewidth]{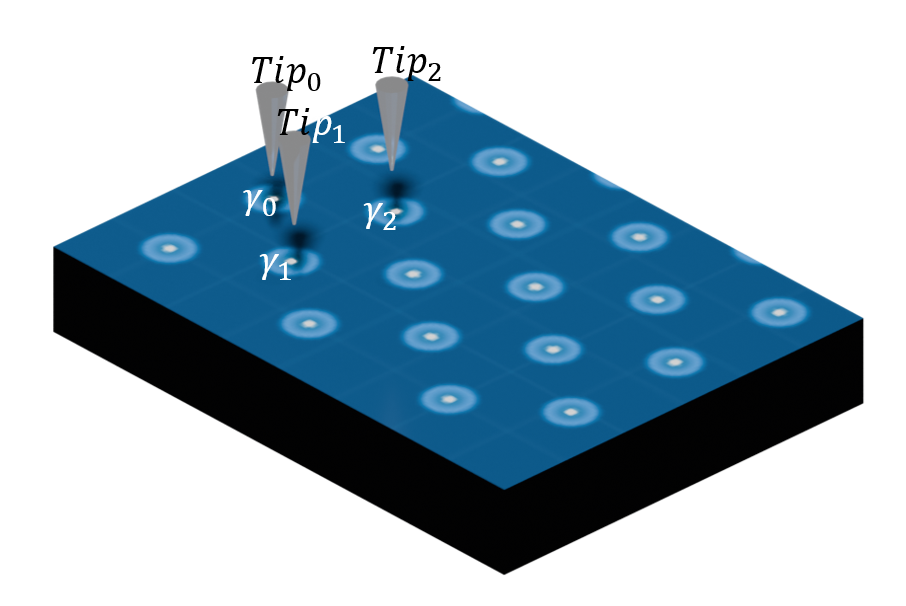}
\put(0,70){\textbf{(b)}}
\end{overpic}
\label{fig:setup_vortex}
}
\caption{(a) Diagram of two topological superconducting nanowires (gray bars) on a superconducting island with Coulomb blockade. 
At the end of the superconducting nanowires, there reside four Majorana zero modes. 
While all four MZMs are present in the system, our study focuses primarily on three Majorana zero modes ($\gamma_0$, $\gamma_1$, $\gamma_2$), each of them is connected to a lead ($V_0$, $V_1$, $V_2$).
Reference arms ($t_{R,10}$ by red and black dashed line, $t_{R,20}$ by yellow and black dashed line, $t_{R,21}$ by blue and black dashed line) provide interference paths between cotunneling paths $\mathcal{O}_{01}$, $\mathcal{O}_{02}$, and $\mathcal{O}_{12}$. The coupling strength of the reference arms can be manipulated by the corresponding gate. 
Circular dashed regions ($\Phi_{01}$, $\Phi_{12}$, $\Phi_{02}$) indicate magnetic flux that controls the phase differences between each reference arm path and its corresponding cotunneling path.
(b) 3D visualization of the possible experimental implementation, showing the 2D topological superconducting surface with localized MZM vortices ($\gamma_0$, $\gamma_1$, $\gamma_2$) coupled to scanning tunneling microscope tips (Tip$_0$, Tip$_1$, Tip$_2$) that serve as leads.}
\label{fig:setup}
\end{figure}

In this section, we introduce a system consisting of three spatially separated Majorana zero modes ($\gamma_0, \gamma_1, \gamma_2$), each coupled to an individual lead. 
Due to their spatial separation, any direct coupling between the MZMs can be safely ignored in our model.
These MZMs reside on a superconducting island under Coulomb blockade conditions. 
To enable controlled interactions between these MZMs, we incorporate reference arms that provide interference paths between the leads. 
The above setup can be experimentally realized in various topological superconducting platforms, 
such as a pair of one-dimensional topological superconducting nanowires or a two-dimensional topological superconducting surface, 
as illustrated in Fig.~\ref{fig:setup}.

The total Hamiltonian of this system comprises four distinct components:

\begin{equation}
H = H_{\mathrm{leads}} + H_C + H_T + H_{\mathrm{ref}}.
\label{total Hamiltonian}
\end{equation}

The leads are modeled as reservoirs of non-interacting fermions: $
H_{\mathrm{leads}} = \sum_{\alpha=0}^{2} \sum_{k} \xi_{k\alpha} c_{\alpha, k}^{\dagger} c_{\alpha, k},
$
where $\xi_{k\alpha} = \frac{p_\alpha^2}{2m} - V_\alpha$ is the energy of the lead $\alpha$ with bias voltage $V_\alpha$.  $c_{\alpha,k}^{\dagger}$ and $c_{\alpha,k}$ are the creation and annihilation operators for electrons in lead $\alpha$ with momentum $k$, respectively.

The Coulomb charging energy of the superconducting island can be expressed as \cite{dittrich1998quantum,fu2010Electron,zhang2019Next,manousakis2020Weak}:
\begin{equation}
H_C = \frac{(\hat{n}e - Q_G)^2}{2C},
\label{eq:H_C quadratic}
\end{equation}
where $\hat{n}$ is the number operator for the excess charge on the island, $Q_G$ is the gate charge, and $C$ is the capacitance of the island. By tuning the gate charge $Q_G$, the charging energy can be effectively controlled. 

The tunneling term $H_T$ couples each lead $\alpha$ to the corresponding MZM $\gamma_\alpha$:

\begin{equation}
H_T = \sum_{\alpha=0}^{2} \sum_{k} (\lambda_{\alpha} \gamma_\alpha c_{\alpha,k} + \lambda_{\alpha}^{*} c_{\alpha,k}^{\dagger} \gamma_\alpha),
\end{equation}
where $\lambda_{\alpha}$ is the tunneling amplitude.

Finally, the reference arms provide direct tunneling paths between different leads:

\begin{equation}
H_{\mathrm{ref}} = \sum_{\alpha'=0}^{1} \sum_{\substack{\alpha=1 \\ \alpha > \alpha'}}^{2} \sum_{k, k'} (t_{R,\alpha \alpha'} c_{\alpha k}^{\dagger} c_{\alpha' k'} + \text{H.c.}),
\end{equation}
where $t_{R,\alpha \alpha'}$ is the amplitude for direct electron transfer between leads $\alpha$ and $\alpha'$ via the reference arm (assuming weak $k, k'$ dependence).

As evident from the Hamiltonians presented above, our theoretical framework explicitly incorporates only three MZMs ($\gamma_0,\,\gamma_1,\,\gamma_2$). Subsequently, we demonstrate that cotunneling-mediated braiding operations are entirely confined within this three-MZM subsystem. While a fourth MZM is typically present in conventional device architectures [see Fig.~\ref{fig:setup_2d}], since MZMs always appear in pair, it remains irrelevant to the cotunneling processes described herein. Rather, the fourth MZM serves to facilitate protected qubit encoding within a sector of fixed total fermion parity\cite{sarma2015majorana,aasen2016Milestones}, as will be shown below.

\subsection{Interference and Effective MZM Coupling} \label{subsec:interference_coupling}
The effective Hamiltonian of Eqs.~(3)-(5) is:
\begin{equation} 
  H_I = \sum_{\alpha'=0}^{1} \sum_{\substack{\alpha=1 \\ \alpha > \alpha'}}^{2} \sum_{k, k'}(t_{R,\alpha \alpha'} + \mathcal{O}_{\alpha\alpha'})(c_{\alpha,k}^\dagger c_{\alpha',k'} + \text{H.c.}).
  \label{eq:HI_main total tunneling}
\end{equation}
The derivation can be found in Appendix~\ref{app:deriv cotunneling schrieffer-wolff}. Physically, this effective Hamiltonian \eq{eq:HI_main total tunneling} is responsible for electron transfer between leads $\alpha'$ and $\alpha$ that involves both the direct reference arm path and the MZM-mediated cotunneling path.
The operator $\mathcal{O}_{\alpha \alpha'}= t_{M,\alpha \alpha'} i \gamma_\alpha \gamma_{\alpha'}$ represents the MZM-mediated cotunneling process between leads $\alpha$ and $\alpha'$, where $t_{M,\alpha \alpha'} = i\frac{2 \lambda_{\alpha}^* \lambda_{\alpha'}}{E_c}$ is the effective cotunneling amplitude. 

When this interaction $H_I$ couples the MZM system to the itinerant electron modes of the fermionic reservoirs, it leads to both dissipation and a coherent shift in the MZM energies, 
known as the Lamb shift \cite{Breuer2002theory,Wiseman2009Quantum}, $H_{LS}$. 
This coherent coupling arises specifically from the 
interference between the two pathways ($t_{R,\alpha \alpha'}$ and $\mathcal{O}_{\alpha \alpha'}$). 
A full derivation using master equations, outlined in Appendix~\ref{app:deriv lindblad master equation}, 
averages out the lead degrees of freedom to obtain the effective Hamiltonian governing the coherent MZM dynamics as:
\begin{align}
  H_{LS} &\approx \sum_{\alpha'=0}^{1} \sum_{\substack{\alpha=1 \\ \alpha > \alpha'}}^{2}\Delta_{\alpha \alpha'} i\gamma_{\alpha}\gamma_{\alpha'}.
  \label{eq:H_LS_main corresponding appendix}
\end{align}
The coupling strength $\Delta_{\alpha \alpha'}$:
\begin{equation} \label{eq:Delta_coupling_via_tunneling}
\begin{aligned}
\Delta_{\alpha \alpha'} &= -2\nu^{2}\Lambda\left|t_{R,\alpha \alpha'}^{*}t_{M,\alpha\alpha'}\right|\cos\left(\frac{2\pi\Phi_{\alpha\alpha'}}{\Phi_{0}}\right),
\end{aligned}
\end{equation}
where $\nu$ is the density of states in the leads (assumed constant near the Fermi level) and $\Lambda$ is a bandwidth cutoff. 
Here $\Phi_{\alpha\alpha'}$ is the phase difference between the direct tunneling amplitude ($t_{R,\alpha \alpha'}$) 
and the MZM-mediated cotunneling amplitude ($t_{M,\alpha\alpha'}$),
and $\Phi_0 = h/e$ is the flux quantum. The tunability of the magnitude 
and sign of $\Delta_{\alpha \alpha'}$ via the Aharonov-Bohm phase ($2\pi \Phi_{\alpha\alpha'}/ \Phi_0$) 
is the central mechanism enabling the braiding protocols described later, 
particularly in cancelling out the dynamic phases.

The interference term $\operatorname{Re}(t_{R,\alpha \alpha'}^* t_{M,\alpha\alpha'})$ not only generates the coherent coupling $\Delta_{\alpha \alpha'}$ but also contributes to the electrical conductance 
$G_{\alpha \alpha'} \approx (2e^2/h)2\pi^2\nu^2(|t_{R,\alpha \alpha'}|^2 + |t_{M,\alpha\alpha'}|^2 + 2\operatorname{Re}(t_{R,\alpha \alpha'}^* t_{M,\alpha\alpha'})i\gamma_{\alpha}\gamma_{\alpha'})$, 
with experimental observation of $h/e$-periodic oscillations confirming this interference \cite{whiticar2020coherent}. 
Comparing the Lamb shift term $|(H_{LS})_{\alpha \alpha'}| \approx 2\nu^2\Lambda\operatorname{Re}(t_{R,\alpha \alpha'}^* t_{M,\alpha\alpha'})$ with quantum interference contribution to conductance $|\Delta G_{\alpha \alpha'}|=(2e^{2}/h)4\pi^{2}\nu^{2}\operatorname{Re}(t_{R,\alpha \alpha'}^{*}t_{M,\alpha\alpha'})$, 
we estimate the MZM coupling strength as $|\Delta_{\alpha \alpha'}| \sim \Lambda|\Delta G_{\alpha \alpha'}|/((2e^2/h)2\pi^2)$. 
Using typical measurement results for the conductance oscillation amplitude 
from Ref.~\cite{whiticar2020coherent}, $ |\Delta G_{\alpha \alpha'}|/(2e^2/h) \sim 7.5\times10^{-3}$ and $\Lambda \sim 1$ eV for InAs, 
we obtain $|\Delta_{\alpha \alpha'}| \approx 0.38$ meV, corresponding to an angular frequency $\omega_{\Delta} = |\Delta_{\alpha \alpha'}|/\hbar \approx 92$ GHz, suggesting experimentally feasible coherent braiding operations within typical MZM coherence time.

This correspondence between cotunneling coherent coupling and quantum interference contribution to conductance also provides a handy tool to measure, 
calibrate, and fine-tune the coherent coupling strength and sign,
which provides us with certain convenience to implement the braiding operations.
Furthermore, the conductance measurement itself can probe the qubit state, which can be integrated into the 
quantum computation process naturally. 
In the following section, we will discuss how to manipulate the cotunneling coherent coupling to implement 
braiding operations to form Clifford gates and implement non-Clifford gates, 
demonstrating that this platform together with the aforementioned measurement convenience can be highly effective for forming a complete topological computing platform.

\section{Cotunneling-Assisted Braiding Protocol} \label{sec:braiding}

\subsection{Theoretical Framework and Cancellation of Dynamic Phase }

\begin{figure*}[t!]
\centering
\subfloat[]{
\begin{overpic}[width=1\textwidth,trim=160 515 80 210,clip]{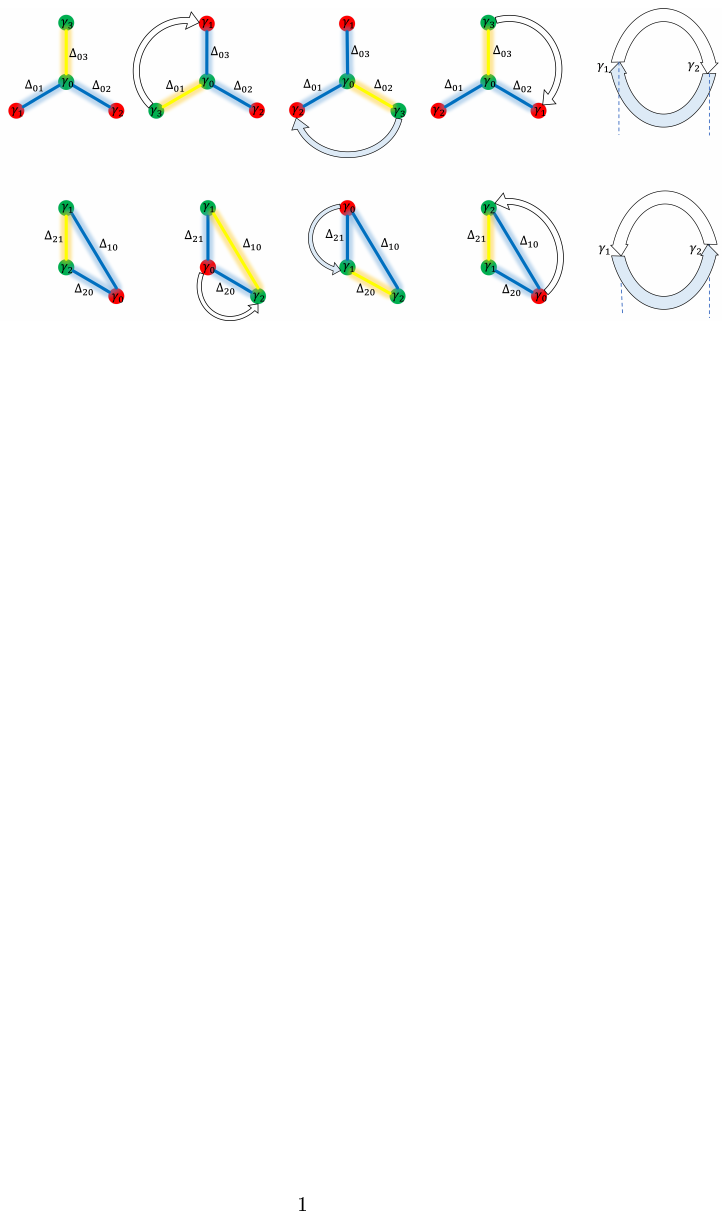}
\put(0,17){\textbf{(a)}}
\end{overpic}
\label{fig:braiding_trijunction}
}
\\
\subfloat[]{
\begin{overpic}[width=1\textwidth,trim=160 595 80 120,clip]{./figs/two_junction_braiding.pdf}
\put(0,19){\textbf{(b)}}
\end{overpic}
\label{fig:braiding_yjunction}
}
\caption{Comparison of two MZM braiding schemes, shown in (a) and (b). 
From left to right, the sequence shows the adiabatic braiding steps, 
the yellow bonds represent the only non-zero coupling between the MZMs at each step,
and arrows represent the motion of the MZMs.
The rightmost diagram shows the topologically equivalent braiding trajectory.
The dashed lines in both panels represent the branch cuts associated with each MZM.
(a) Braiding in our proposed triangle junction with counterclockwise exchange. The sequence illustrates how $\gamma_1$ crosses the branch cut of $\gamma_2$, resulting in the transformation $\gamma_1 \rightarrow -\gamma_2$ and $\gamma_2 \rightarrow \gamma_1$.
(b) Braiding in a Y-junction configuration with clockwise exchange, where $\gamma_2$ crosses $\gamma_1$'s branch cut, resulting in the transformation $\gamma_1 \rightarrow \gamma_2$ and $\gamma_2 \rightarrow -\gamma_1$. 
}
\label{fig:braiding_comparison}
\end{figure*}

Our proposed braiding scheme leverages the controllable, interference-induced coupling between MZMs derived in Sec.~\ref{sec:hamiltonian}. 
The fundamental concept involves utilizing the control gate to modulate $t_{R,\alpha \alpha'}$, thereby adiabatically manipulating the couplings $\Delta_{\alpha \alpha'}$ in \eq{eq:Delta_coupling_via_tunneling}. This controlled manipulation guides the system's ground state through a precisely defined unitary transformation corresponding to a MZM exchange. 
As illustrated in Fig.~\ref{fig:braiding_trijunction}, to execute a braiding between $\gamma_1$ and $\gamma_2$, one implements a sequence where the dominant coupling term rotates through the cycle 
$\Delta_{21} \rightarrow \Delta_{10} \rightarrow \Delta_{20} \rightarrow \Delta_{21}$, where our naming of $\Delta_{\alpha \alpha'}$ is fixed to correspond to the initial arranged positions of each MZM, as shown in Fig.~\ref{fig:braiding_comparison}, despite that during the braiding certain MZMs have been exchanged
. 
When performed adiabatically over a time interval $T_{\mathrm{braid}}$, this evolution theoretically produces a geometric phase transformation corresponding to the braid operator
 $B_{12} = \exp(\frac{\pi}{4} \gamma_1 \gamma_2)$, as demonstrated in Appendix~\ref{app:mzm_braiding}.

However, an important distinction arises between our proposed triangle junction configuration and the conventional Y-junction shown in Fig.~\ref{fig:braiding_comparison}. 
In the Y-junction, ideally two MZMs undergoing braiding remain decoupled and maintain zero energy throughout the process\cite{vanHeck2012Coulomb,Oppen2017Topological}. 
In contrast, within our triangle junction braiding scheme,
the MZMs being braided experience a coupling throughout the braiding and their eigenenergies are not zero. 
This non-zero eigenenergy inevitably leads to the accumulation of a dynamic phase $U_{\mathrm{dyn}} = \exp(-i \int_0^{T_{\mathrm{braid}}} E_i(t) dt)$, where $E_i(t)$ represents the instantaneous $i^{\mathrm{th}}$ state energy. 
Such dynamic phase accumulation is generally uncontrollable and undermines the topological protection that makes MZM braiding valuable for quantum computation\cite{karzig2017scalable,Clarke2017Probability,Coopmans2021Dynamical}.

In fact, even in the Y-junction configuration, 
real physical implementations inevitably introduce finite-size effects that lead to non-zero MZM couplings\cite{Cheng2009Splitting,Cayao2017Majorana,prada2020Andreev}. 
These couplings, though typically small, still result in dynamic phase accumulation during braiding operations\cite{Cheng2011Nonadiabatic,Clarke2017Probability,harper2019Majorana}, 
presenting a similar challenge as in our triangle junction scheme. 
This observation suggests that the dynamic phase cancellation technique to be introduced below 
could have broader implications beyond the triangle junction architecture.

To eliminate the dynamic phase accumulation in our triangle braiding scheme, we employ a spin echo-like technique, enabled by the flux tunability of the MZM couplings [see \eq{eq:Delta_coupling_via_tunneling}]. As established earlier, applying a half flux quantum $\Phi_0/2$ through the interference loops reverses the sign of the reference arm tunneling amplitudes $t_{R,\alpha \alpha'}$, consequently inverting the sign of the effective MZM couplings: $\Delta_{\alpha \alpha'} \rightarrow -\Delta_{\alpha \alpha'}$.

To illustrate the dynamic phase cancellation strategy, we consider the NOT gate as an example first. Prior to proceeding, we fix the qubit encoding used in this and the following sections: the single qubit is encoded in the parity of $\gamma_1$ and $\gamma_2$, in this way the Pauli operators acting on this qubit is $\sigma_x = i\gamma_1\gamma_0,\quad \sigma_y = i\gamma_2\gamma_0,\quad \sigma_z = i\gamma_2\gamma_1$.  
The NOT gate requires two identical braiding operations:
\begin{equation} 
\begin{aligned}
B_{10}^{2} &= \dfrac{1+\gamma_{1}\gamma_{0}}{\sqrt{2}} \cdot \dfrac{1+\gamma_{1}\gamma_{0}}{\sqrt{2}} \\
&= -i\sigma_{x}, \\
\end{aligned}
\label{eq:NOT_gate_B10_squared}
\end{equation}
where the braiding operation $B_{10} = \exp(\frac{\pi}{4} \gamma_1 \gamma_0)$.

The echo protocol can be straightforwardly applied:

\begin{protocol} \label{protocol:simplified}
Simplified Protocol for the NOT gate
\begin{enumerate}
\item Perform a complete braiding operation $B_{10}$ over a duration $T_{\mathrm{braid}}$, accumulating both dynamic phase $U_{\mathrm{dyn},1}$ and geometric phase $U_{\mathrm{geom}}$.
\item Apply a $\pi$-flux ($\Phi \rightarrow \Phi + \Phi_0/2$) to invert the Hamiltonian.
\item Repeat the identical braiding operation $B_{10}$ over another duration $T_{\mathrm{braid}}$, Due to step 2 now $H \rightarrow -H$, the dynamic phase accumulation follows $U_{\mathrm{dyn},2} = U_{\mathrm{dyn},1}^{\dagger}$, while the geometric phase contribution remains identical.
\end{enumerate}
\end{protocol}

The braiding operation $B_{10}$ is achieved by controlling the reference arm tunneling amplitudes 
$t_{R,\alpha\alpha'}$ to drive the system through the following sequence of coherent 
couplings: $\Delta_{10} \rightarrow \Delta_{21} \rightarrow \Delta_{20} \rightarrow \Delta_{10}$,
based on \eq{eq:Delta_coupling_via_tunneling}.
In the subsequent simulation sections, we will provide detailed implementation methods and numerical results demonstrating this process.

However, the implementation of quantum gates with asymmetric braiding sequences, such as the Hadamard gate, presents additional challenges. Consider the Hadamard gate construction:

\begin{align} \label{eq:hadamard_gate}
B_{10}B_{21}B_{10} &= \dfrac{1+\gamma_{1}\gamma_{0}}{\sqrt{2}}\dfrac{1+\gamma_{2}\gamma_{1}}{\sqrt{2}}\dfrac{1+\gamma_{1}\gamma_{0}}{\sqrt{2}} \notag\\
&= \dfrac{-i}{\sqrt{2}}\left(\sigma_{x}+\sigma_{z}\right) = H.
\end{align}

In this sequence, the braiding operation $B_{21}$ occurs only once, precluding the direct application of the echo protocol described for symmetric sequences. 

To address this challenge, we leverage the tunability of the effective coupling $\Delta_{\alpha\alpha'}$ established in \eq{eq:Delta_coupling_via_tunneling} to implement braiding-like operations 
\begin{align}
B_{ij}(\theta) := \exp\!\big[(\theta/2)\,\gamma_i\gamma_j\big]
\label{eq:theta_braiding_like_sequence}
\end{align}
for arbitrary values of $\theta$ through a modified sequence of $\Delta_{\alpha\alpha'}$ as described in \eq{eq:theta_braiding_sequence}. This capability, illustrated in Fig.~\ref{fig:braiding_sequence_bloch} in Appendix~\ref{app:mzm_braiding}, enables us to decompose any exchange operation into smaller segments. In particular, the standard exchange in \eq{eq:hadamard_gate} can be decomposed into $B_{21}:=B_{21}(\tfrac{\pi}{2})=B_{21}(\tfrac{\pi}{4})B_{12}(\tfrac{\pi}{4})$. In general, we propose:
\begin{protocol} \label{protocol:standard}
Standard Protocol for an exchange
\begin{enumerate}
\item For each exchange operation $B_{\alpha\beta}$ in the sequence, decompose it into two operations: $B_{\alpha\beta} = B_{\alpha\beta}(\pi/4)\,B_{\alpha\beta}(\pi/4)$
\item For a single exchange, apply an echo protocol: 
   \begin{enumerate}
   \item Perform $B_{\alpha\beta}(\pi/4)$ over duration $T_{\mathrm{half}}$, accumulating $U_{\mathrm{dyn},1}$ and $U_{\mathrm{geom}}^{1/2}$
   \item Apply a $\pi$-flux to invert the Hamiltonian
   \item Perform the same operation again, accumulating $U_{\mathrm{dyn},1}^{\dagger}$ and another $U_{\mathrm{geom}}^{1/2}$
   \end{enumerate}
\item Combine these echo-protected operations to form the complete gate
\end{enumerate}
\end{protocol}

This strategy enables the implementation of any Clifford gate through a sequence of echo-protected operations, preserving topological protection despite the presence of non-zero energy splittings. 
In the following, we present numerical evidence validating our framework.

\subsection{Dynamic Phase Cancellation: Simulation via Lindblad Master Equation} \label{subsec:dynamic_phase_sim}

\begin{figure*}
  \centering
  \subfloat[]{\begin{overpic}[width=0.47\textwidth]{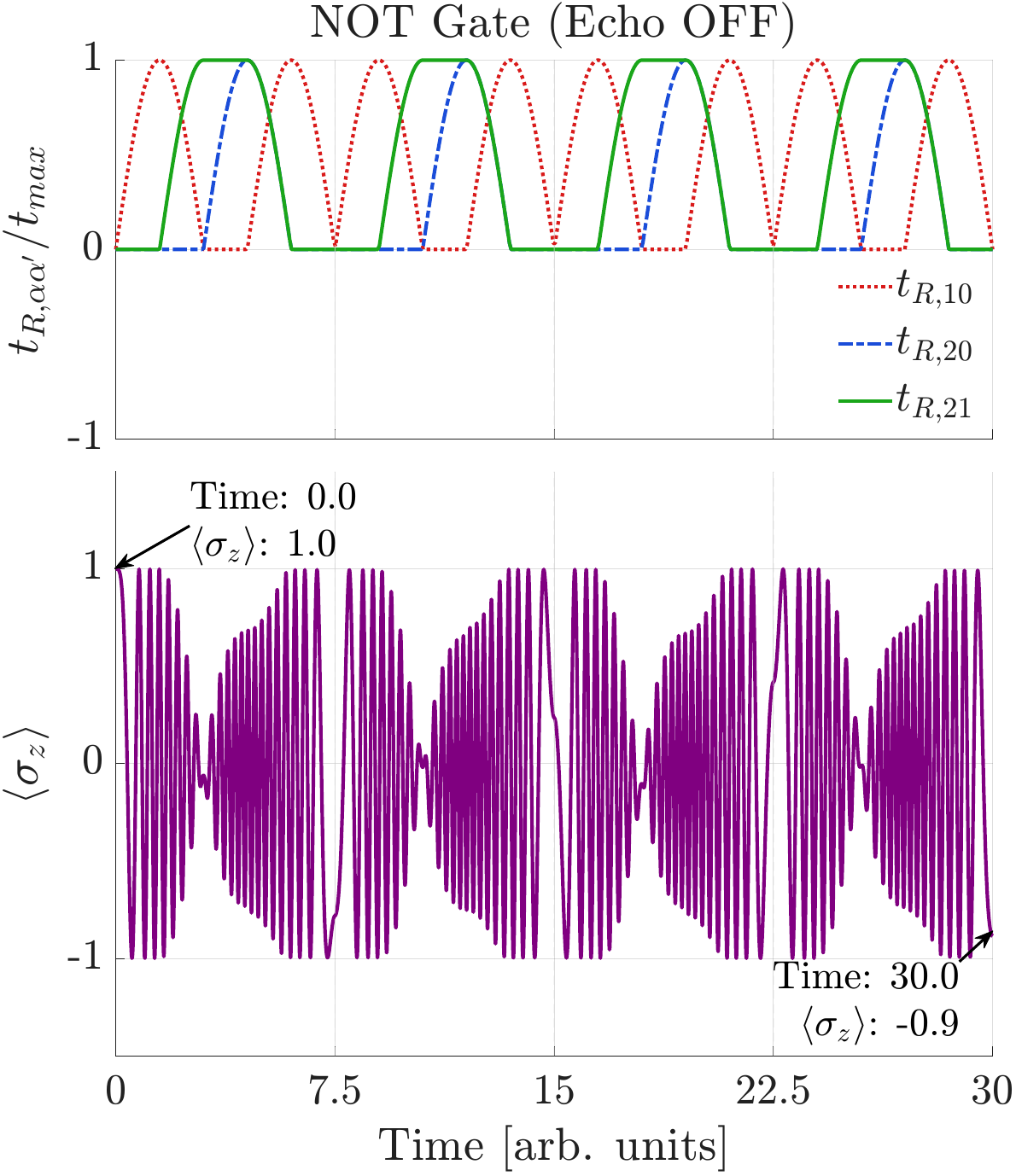}\put(0,97){\textbf{(a)}}\end{overpic}\label{fig:notgate_no_echo_a}}
  \hspace{0.03\textwidth}
  \subfloat[]{\begin{overpic}[width=0.47\textwidth]{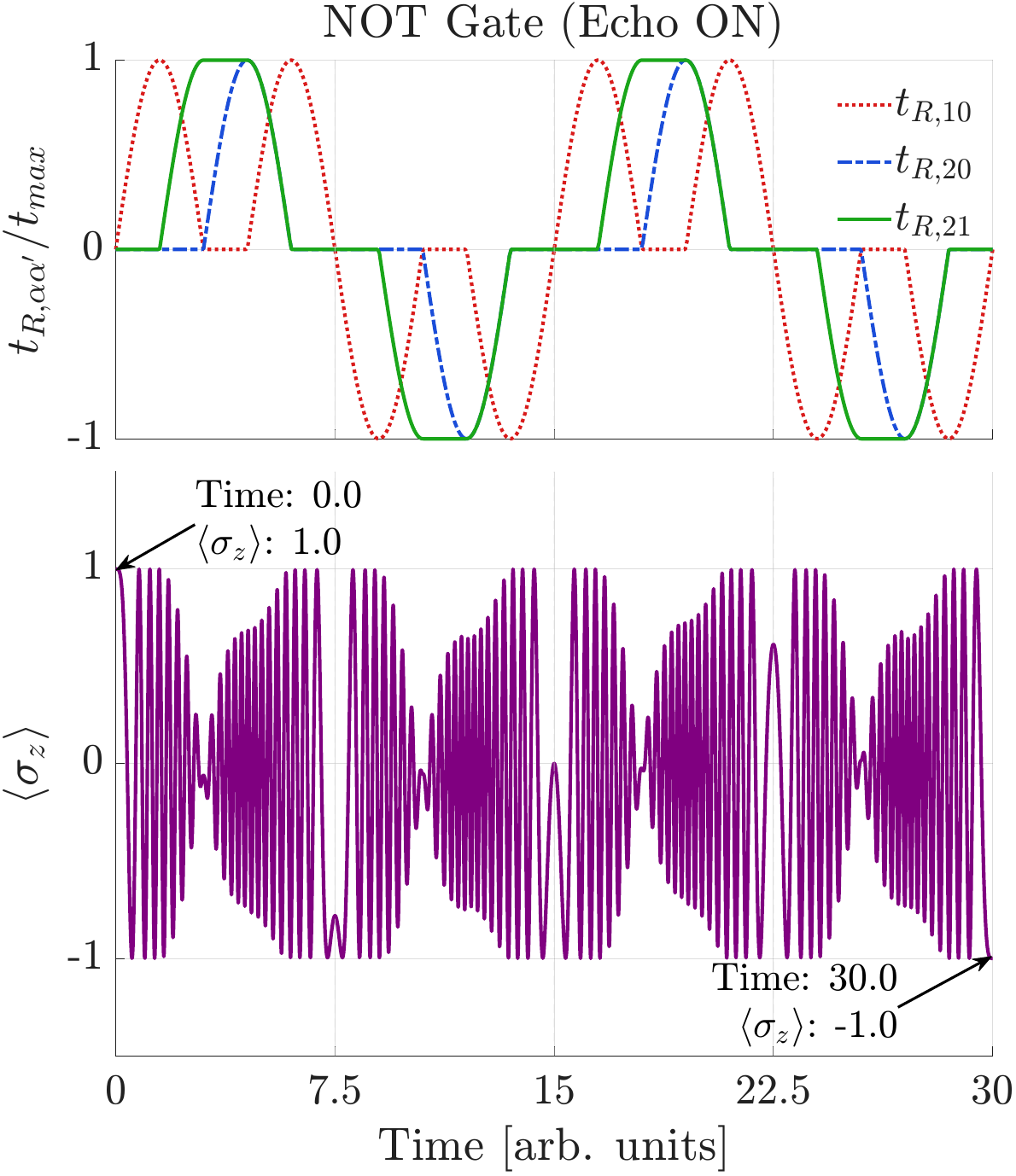}\put(0,97){\textbf{(b)}}\end{overpic}\label{fig:notgate_with_echo_b}}
  
  \subfloat[]{\begin{overpic}[width=0.47\textwidth]{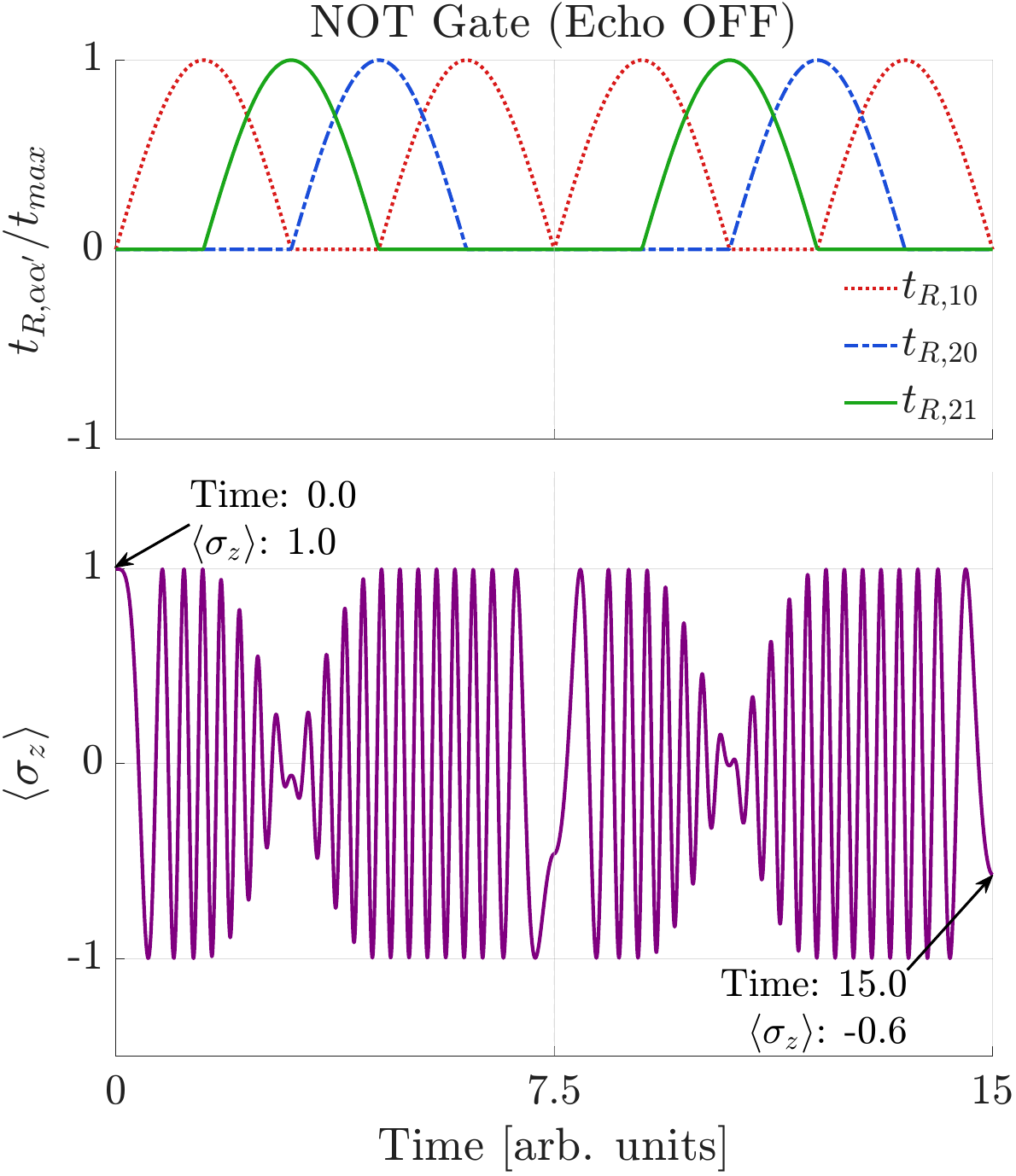}\put(0,97){\textbf{(c)}}\end{overpic}\label{fig:notgate_shortcut_no_echo_c}}
  \hspace{0.03\textwidth}
  \subfloat[]{\begin{overpic}[width=0.47\textwidth]{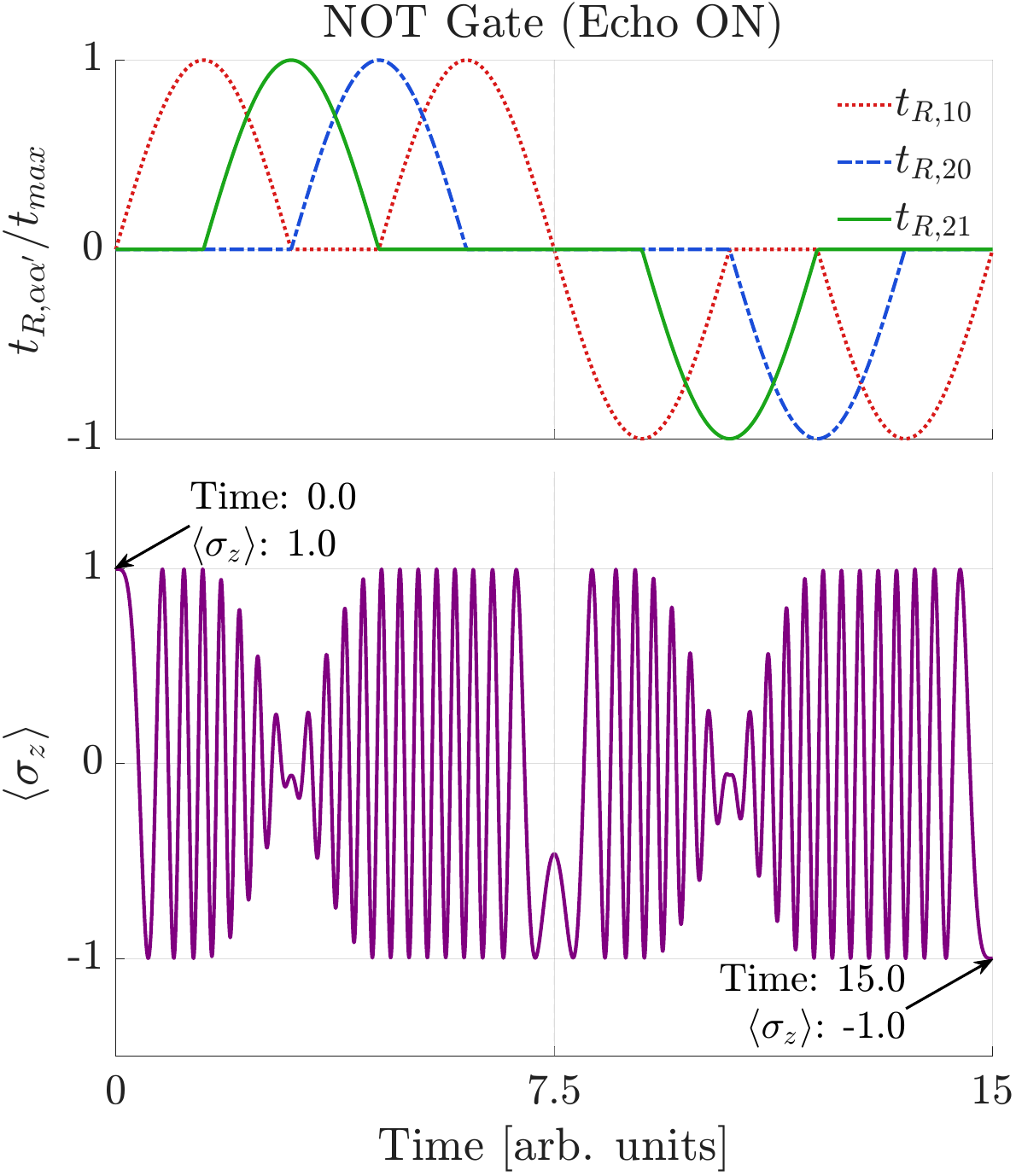}\put(0,97){\textbf{(d)}}\end{overpic}\label{fig:notgate_shortcut_with_echo_d}}
  \caption{\label{fig:not_gate_simulation_long} NOT gate simulations: parameters are $\nu=1.0$, $\Lambda=10.0$, temperature $T=1.0\times10^{-10}$. 
  (a) Standard method without echo protection, (b) standard method with echo protection, (c) simplified method without echo protection, and (d) simplified method with echo protection. For each subplot: The upper panel shows the modulation functions of reference arm tunneling parameters
   $t_{R,10}$, $t_{R,20}$, and $t_{R,21}$ scaled by their maximum value $t_{\text{max}} = 0.5$ used in this simulation;
   the lower panel shows the corresponding expectation value of $\sigma_z$. Note we also annotate the expected value of $\sigma_z$ at the start and end of the braiding operation. Expectation value of $\sigma_{z}$ evolving from $1.0$ to $-1.0$
   represents a high-fidelity NOT gate operation. }
\end{figure*}

To verify our theoretical framework, we performed extensive numerical simulations of the braiding process in our triangle junction scheme. Our simulations employ the Lindblad master equation approach:
\begin{equation}
\frac{d}{dt}\rho_{S}(t)=-i\left[H_{LS},\rho_{S}(t)\right]+\mathcal{D}\left(\rho_{S}(t)\right),
\end{equation}
where $H_{LS}$ is the Lamb shift Hamiltonian derived in Section~\ref{sec:hamiltonian}, and the dissipation term $\mathcal{D}(\rho_{S})$ is given by:
\begin{equation}
\begin{aligned}
\mathcal{D}\left(\rho_{S}\right) = \sum_{\omega}\sum_{\alpha\alpha^{\prime}}&\Gamma_{\alpha\alpha^{\prime}}(\omega) \Big(A_{\alpha^{\prime}\alpha}(\omega)\rho_{S}A_{\alpha\alpha^{\prime}}(-\omega) \\
&-\frac{1}{2}\left\{ A_{\alpha\alpha^{\prime}}(-\omega)A_{\alpha^{\prime}\alpha}(\omega),\rho_{S}\right\} \Big).
\end{aligned}
\end{equation}
The operators $A_{\alpha\alpha'}$ appearing in this equation are defined as $A_{\alpha\alpha'} = t_{R,\alpha \alpha'} + \mathcal{O}_{\alpha\alpha'}$, 
combining the direct tunneling and cotunneling pathways. 
For the unbiased leads we considered here, the dissipation rate $\Gamma_{\alpha\alpha'}$ is given by $\Gamma_{\alpha\alpha'} = 2\pi \nu^{2}T$, where $\nu$ is the density of states of the lead and $T$ is the lead temperature.
This approach captures both the coherent dynamics and the dissipation through the fermionic leads. 
A detailed derivation of this master equation can be found in Appendix~\ref{app:deriv lindblad master equation}.

For the NOT gate, as demonstrated in the last subsection
there are two implementations: 
The first approach, referred to as the ``standard" method, employs Protocol~\ref{protocol:standard} where each individual braiding operation $B_{10}$ is decomposed into operations $B_{10}(\pi/4)$ with echo protection. Since the NOT gate requires two braiding operations $B_{10}^2$, this method applies $\pi$-flux twice.
The second approach, referred to as the ``simplified" method, employs Protocol~\ref{protocol:simplified} directly to the entire NOT gate operation $B_{10}^2$ as a unified process. This method treats the two braiding operations as a single sequence and applies echo protection only once, making it more efficient while achieving the same dynamic phase cancellation.

The standard method, as illustrated in Fig.~\ref{fig:not_gate_simulation_long}(a-b), employs two identical braiding operations $B_{01}$, each with a duration $T_{\mathrm{braid}}$. Each braiding operation is further decomposed into two operations $B_{01}(\pi/4)$ of duration $T_{\mathrm{braid}}/2$. 

To prevent unwanted dynamic phase accumulation during idle periods, the control parameters are initialized and terminated with all reference tunneling parameters set to zero for each operation. As demonstrated in Fig.~\ref{fig:not_gate_simulation_long}(a-b) and supported by \eq{eq:Delta_coupling_via_tunneling}, each operation follows a specific coherent coupling evolution: $\Delta_{10} \rightarrow \Delta_{21} \rightarrow \Delta_{21}+\Delta_{20} \rightarrow \Delta_{10}$. The intermediate step $\Delta_{21}+\Delta_{20}$ is achieved by simultaneously activating $t_{R21}$ and $t_{R20}$ to identical values. 
This precise tuning is facilitated by the correspondence between cotunneling coherent coupling and quantum interference contribution to conductance discussed in Sec.~\ref{subsec:interference_coupling}.

In the absence of echo protection [Fig.~\ref{fig:notgate_no_echo_a}], the operation suffers from distortion due to dynamic phase accumulation. However, the echo implementation [Fig.~\ref{fig:notgate_with_echo_b}] employs a midpoint echo ($t_{R,\alpha\alpha'} \rightarrow -t_{R,\alpha\alpha'}$) that effectively cancels the dynamic phase while preserving the geometric phase. This crucial feature enables the realization of high-fidelity NOT gate operations.

To quantitatively demonstrate the effectiveness of our echo protocol, we examine the final $\sigma_z$ expectation values for both implementations. 
For the standard method, without echo protection [Fig.~\ref{fig:notgate_no_echo_a}], the intended NOT gate only flips the initial $\sigma_z = 1.0$ state to $\sigma_z = -0.9$, indicating significant dynamic phase distortion. 
In contrast, with echo protection [Fig.~\ref{fig:notgate_with_echo_b}], the NOT gate achieves a high-fidelity flip from $\sigma_z = 1.0$ to $\sigma_z = -1.0$. 
Similarly, for the simplified method [Fig.~\ref{fig:not_gate_simulation_long}(c-d)], without echo the intended NOT gate only transforms the initial state from $\sigma_z = 1.0$ to $\sigma_z = -0.6$, while with echo protection it achieves the ideal transformation from $\sigma_z = 1.0$ to $\sigma_z = -1.0$. 
These quantitative results clearly demonstrate that both our standard and simplified methods can eliminate dynamic phase accumulation and provide high-fidelity topological NOT gate operations when combined with the echo protocol.

\begin{figure}[htp!]
\centering
\begin{overpic}[width=1.0\linewidth]{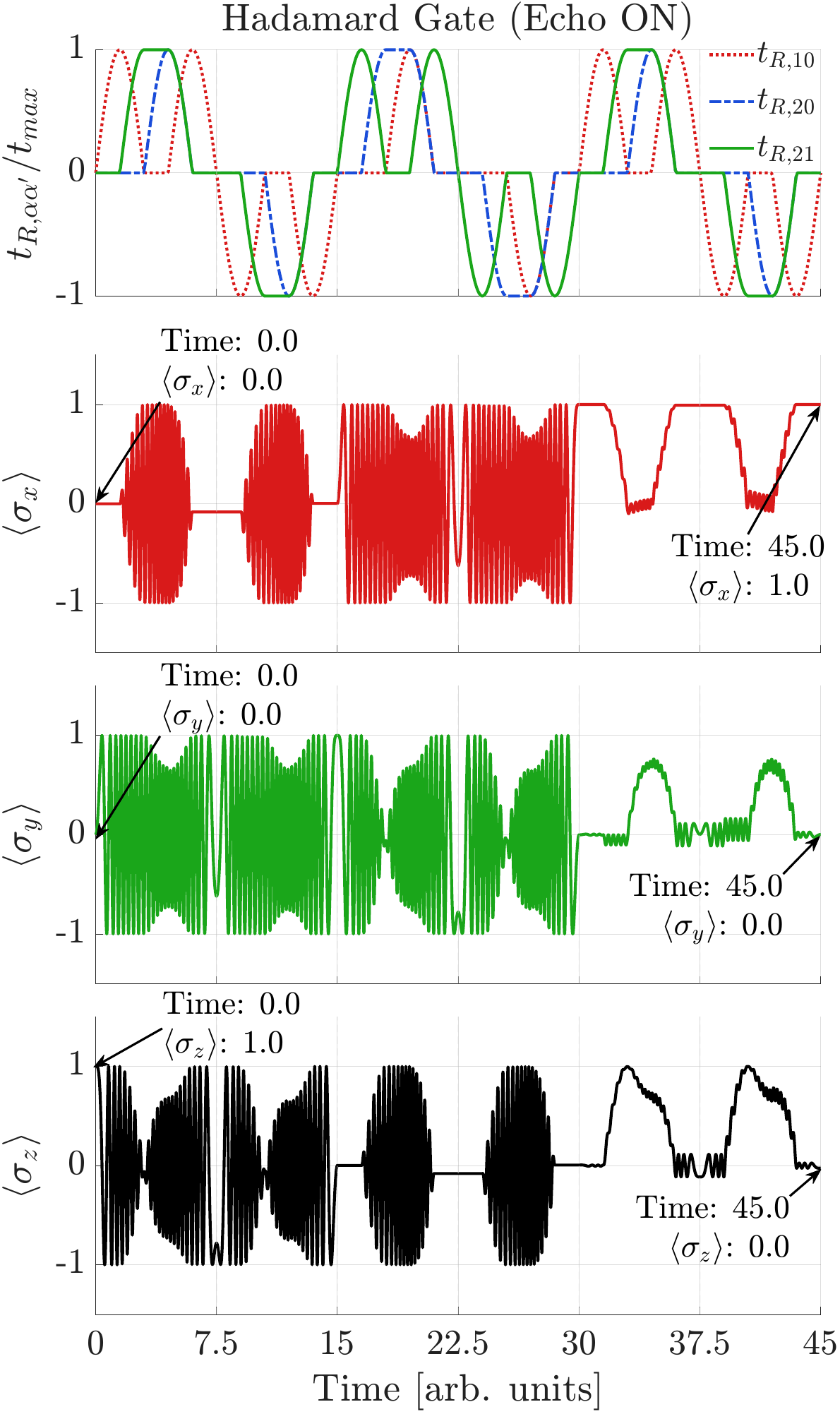}
  \put(0,97){\textbf{(a)}\label{fig:hadamard_params}}
  \put(0,75){\textbf{(b)}\label{fig:hadamard_expectation}}
\end{overpic}
\caption{Hadamard gate simulation with echo protection.   Simulation parameters are identical to those used in Fig.~\ref{fig:not_gate_simulation_long}.
(a) The modulation function of reference arm tunneling parameters $t_{R,10}$, $t_{R,20}$, and $t_{R,21}$ implementing 
the asymmetric braiding sequence $B_{10} \cdot B_{20} \cdot B_{10}$. 
(b) The corresponding expectation values of $\sigma_x$, $\sigma_y$, and $\sigma_z$. 
The initial $\sigma_z = 1$ state is correctly transformed to $\sigma_x = 1$ at the end of the operation, 
demonstrating a successful Hadamard transformation with dynamic phase cancellation. 
}
\label{fig:hadamard_with_echo}
\end{figure}

We also verified our approach with more complex Clifford gates like the Hadamard gate, as 
shown in \eq{eq:hadamard_gate}, this gate operation is a $\pi$ rotation 
around the $\frac{\hat{x}+\hat{z}}{\sqrt{2}}$ axis on the Bloch sphere.
 
Figure~\ref{fig:hadamard_with_echo} shows the simulation results for the Hadamard gate 
with echo protection, demonstrating that an initial state in the $\sigma_z$ basis is correctly 
transformed to the $\sigma_x$ basis, as expected from the Hadamard operation. 
These results further confirm the effectiveness of the dynamic phase cancellation technique across different Clifford gates.

These simulation results demonstrate that our theoretical framework for dynamic phase cancellation works effectively in practice. 
More simulation results exploring different braiding time durations to demonstrate the necessity of sufficient adiabaticity for achieving high-fidelity gate, 
as well as different reference arm $t_{R,\alpha \alpha'}$ modulation functions to demonstrate the generality and robustness of this method, 
can be found in Appendix~\ref{app:additional_simulation}. 

\section{Geometric $T$ Gate Implementation} \label{sec:tgate}

\begin{figure}[!htbp]
  \subfloat[]{\begin{overpic}[width=\linewidth]{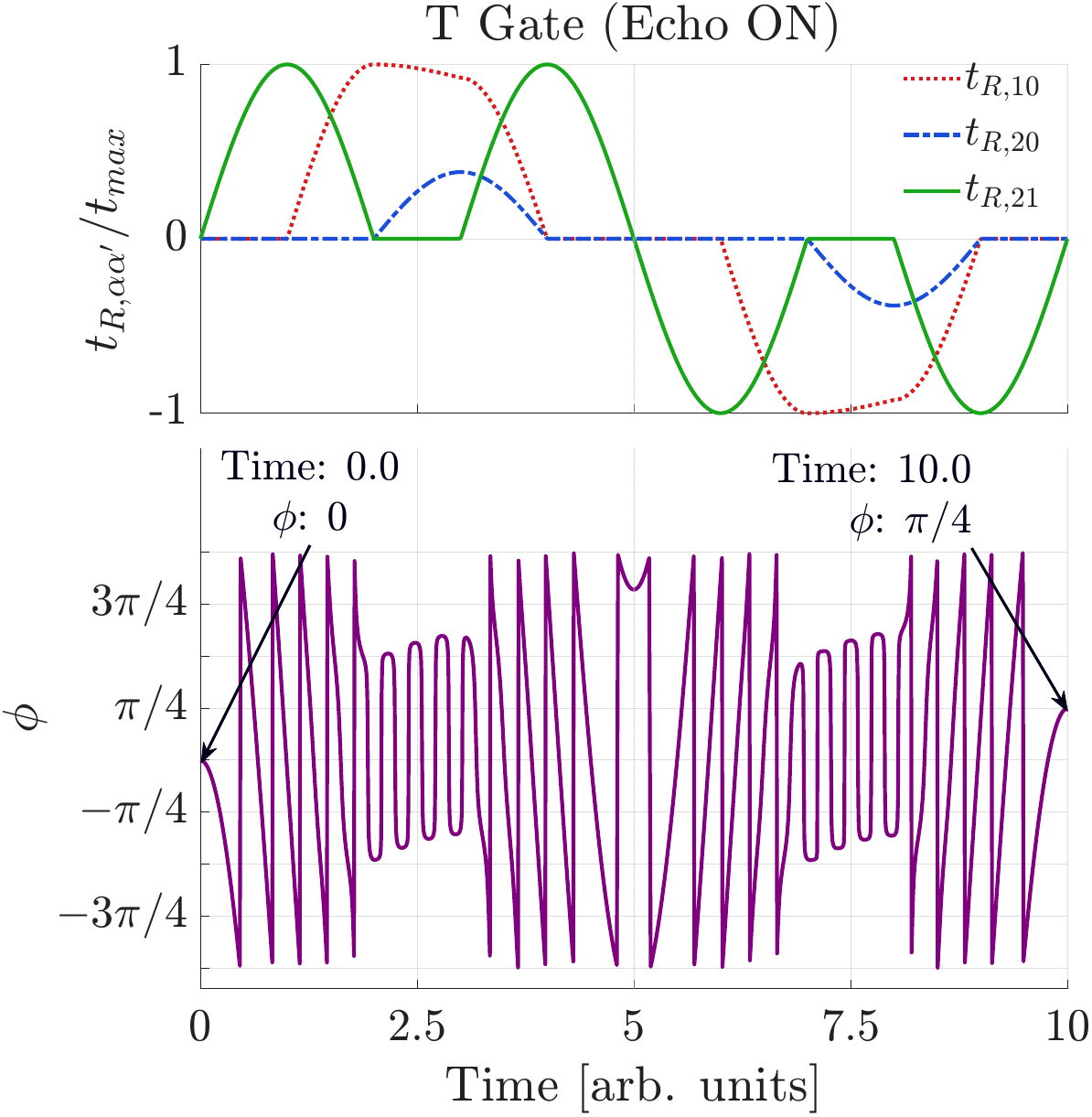}\put(1,97){\textbf{(a)}}\end{overpic}\label{fig:tgate_combined}}\\
  \subfloat[]{\begin{overpic}[width=0.8\linewidth,trim=100 140 100 170,clip]{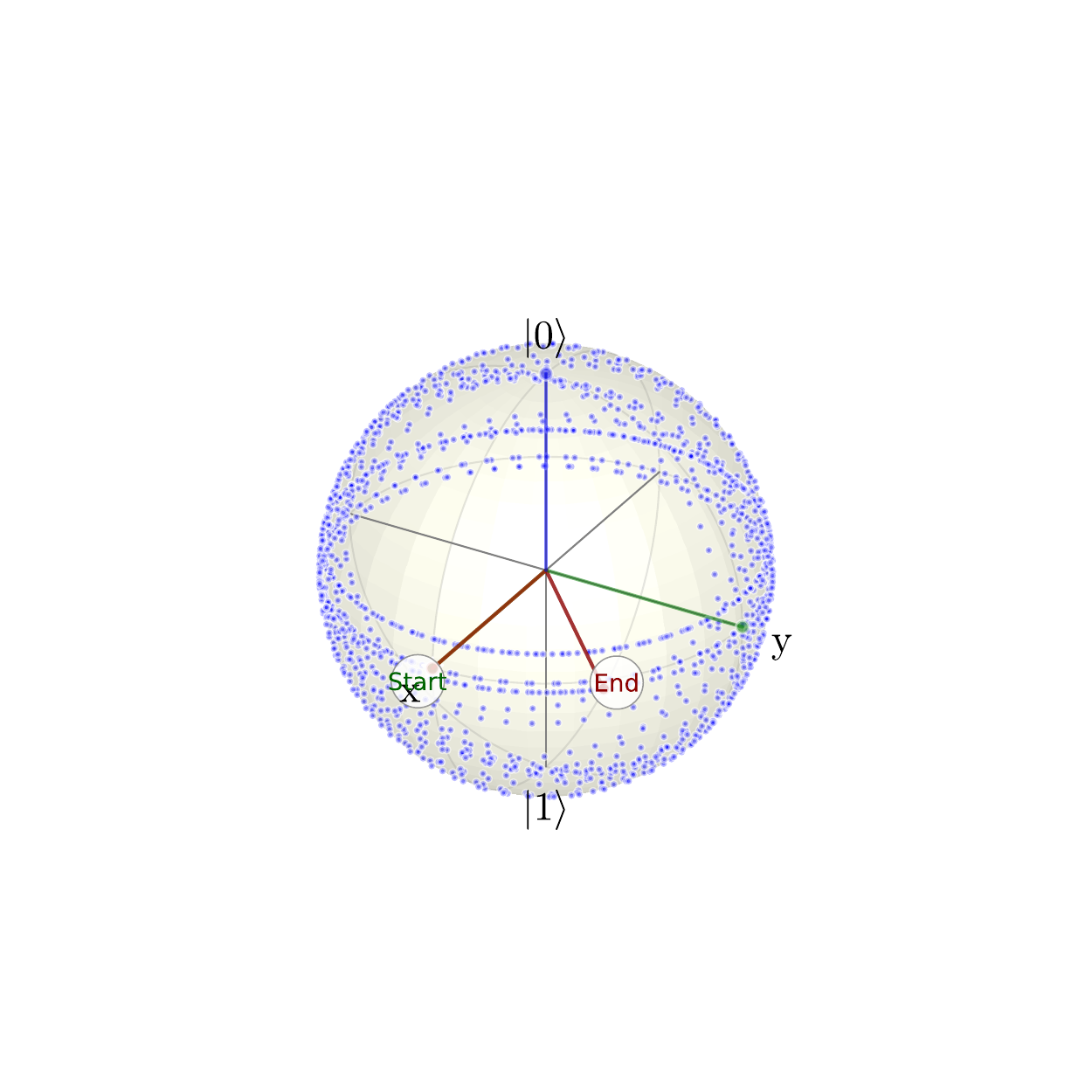}\put(-9,77){\textbf{(b)}}\end{overpic}\label{fig:tgate_bloch}}
  \caption{$T$ gate simulation with echo protection. Simulation parameters are identical to those used in Fig.~\ref{fig:not_gate_simulation_long} except the braiding speed has been adjusted for clearer visualization of the current gate operation.
  (a) The modulation function of reference arm parameters and corresponding relative phase $\phi$ between spin-up and spin-down components of the qubit $|\psi\rangle=\cos \left(\frac{\theta}{2}\right)|\uparrow\rangle+e^{i \phi} \sin \left(\frac{\theta}{2}\right)|\downarrow\rangle$.  The qubit starts from the $\sigma_x = +1$ state which has $\phi=0$ and ends up with $\phi=\pi/4$ as the annotation indicates. 
  (b) Bloch sphere representation showing the state evolution path. Note that the relative phase $\phi$ corresponds to the azimuthal angle of the Bloch vector; the start and end points are annotated by empty circles, and the remaining blue dots represent the intermediate states.
  }
  \label{fig:tgate}
\end{figure}

A significant advantage of our cotunneling-assisted platform is its potential to realize non-Clifford gates, which are essential for universal quantum computation, 
through purely geometric phase mechanisms, thereby extending topological protection beyond the Clifford group.
Braiding operations inherently generate only Clifford gates. 
Conventional approaches introduce non-Clifford gates, like the $T$ gate ($U_T = \text{diag}(1, e^{i\pi/4})$), 
through dynamically controlled interactions \cite{sarma2015majorana, karzig2017scalable,beenakker2020Search}, for example, by bringing two MZMs close together for a specific time duration $t$ to induce an energy splitting $\Delta E$, 
yielding $U = \text{diag}(1, e^{i \Delta E t})$. This reliance on precise timing and energy control makes the non-Clifford gate vulnerable to errors and decoherence, 
undermining the fault tolerance that topological quantum computation aims to achieve.

On the other hand, up to an irrelevant global phase, the $T$ gate can be written as:
\begin{equation}
  U_T=e^{-i\frac{\pi}{8}\sigma_z}=B_{21}(\tfrac{\pi}{4})=B_{21}(\tfrac{\pi}{8})B_{21}(\tfrac{\pi}{8}).
\end{equation}
This scheme offers an alternative by leveraging geometric phases. 
Similar to what we have done in the previous section \ref{sec:braiding}, we can implement the $T$ gate by:
\begin{protocol} \label{protocol:tgate}
$T$ Gate Protocol
\begin{enumerate}
\item Perform $B_{21}(\tfrac{\pi}{8})$.
\item Apply a $\pi$-flux to invert the couplings: $\Delta_{\alpha \alpha'} \rightarrow - \Delta_{\alpha \alpha'}$.
\item Repeat $B_{21}(\tfrac{\pi}{8})$.
\end{enumerate}
\end{protocol}
The coherent coupling $\Delta_{\alpha\alpha'}$ sequence to realize $B_{21}(\tfrac{\pi}{8})$ can be found in Appendix~\ref{app:mzm_braiding}.

Due to the echo, the dynamic phase cancels out. The total accumulated geometric phase shift between the qubit states $|\uparrow\rangle$ and $|\downarrow\rangle$ is $\pi/4$. 
To demonstrate the feasibility of this geometric $T$ gate implementation, we performed numerical simulations using the same master equation approach described earlier. 
Figure~\ref{fig:tgate} shows the simulation results for the $T$ gate with echo protection. 
The system was initialized in a state of $\sigma_x = 1$, and the reference arm tunneling parameters were controlled to implement the operator sequence described above. 
It can be seen in Fig.~\ref{fig:tgate_bloch} that the final state is rotated by $\pi/4$ around the $\hat{z}$ basis, as expected for the $T$ gate operation.

These simulation results confirm that our cotunneling-assisted platform can implement the $T$ gate with high fidelity using purely geometric mechanisms. To verify the generality of our $T$ gate protocol, we also tested it on different initial states as shown in Appendix~\ref{app:tgate_initial_states}. The additional simulation results demonstrate that our implementation achieves high fidelity realization of $U_T=e^{-i\frac{\pi}{8}\sigma_z}$. By combining this non-Clifford operation with the Clifford gates demonstrated earlier, our system provides a complete set of gates for universal quantum computation, all protected by the same dynamic phase cancellation technique.

\section{Conclusion} \label{sec:conclusion}
In summary, we have presented a novel approach to Majorana zero mode braiding that leverages cotunneling processes and interference effects in a 3-MZM system. 
Our scheme offers several key advantages: minimal scheme requiring only three MZMs, tunable coupling through interference effects, and dynamic phase cancellation via flux control. 
Most importantly, we demonstrate that this platform can implement both Clifford gates and non-Clifford gates such as the $T$ gate solely by geometric phases, providing a complete set of gates for universal quantum computation while maintaining topological protection. 
The numerical simulations using quantum master equations validate our theoretical framework and demonstrate the robustness of the protocol.

\begin{acknowledgments}
  This work is financially supported by the National Key R\&D Program of China (Grants No. 2019YFA0308403), the Innovation Program for Quantum Science and Technology (Grant No. 2021ZD0302400), the National Natural Science Foundation of China (Grant No. 12304194), and Shanghai Municipal Science and Technology (Grant No. 24DP2600100).
\end{acknowledgments}

\appendix

\section{Derivation of the MZM-Mediated Cotunneling Operator} \label{app:deriv cotunneling schrieffer-wolff}
As shown in main text Fig.~\ref{fig:setup}, the fourth MZM $\gamma_3$ together with $\gamma_0$, $\gamma_1$, and $\gamma_2$ can form two Dirac fermions with occupation numbers $n_1 = \frac{1}{2}(1 + i\gamma_0\gamma_1)$ and $n_2 = \frac{1}{2}(1 + i\gamma_2\gamma_3)$. Without Coulomb interaction, 
the four states $n_1=0,1$ and $n_2=0,1$ are degenerate at zero energy. The charging energy breaks this degeneracy. 
By adjusting the gate charge $Q_G$ to match the Coulomb valley with the $n_1 + n_2 = 1$ charge configuration, we obtain the following simplified Hamiltonian:

\begin{equation}
  H_C = \begin{cases}
  E_C & \text{if } n_1 = n_2 = 0 \\
  0 & \text{if } n_1 = 1, n_2 = 0 \text{ or } n_1 = 0, n_2 = 1 \\
  E_C & \text{if } n_1 = n_2 = 1
  \end{cases},
  \label{eq:H_C_strong cases}
  \end{equation}
where $E_C = \frac{e^2}{2C}$ comes from \eq{eq:H_C quadratic}.
When the Coulomb blockade is strong, i.e., $E_C$ is the largest energy scale in Hamiltonian \eq{total Hamiltonian},
direct tunneling $H_T$ between a lead and the island is energetically suppressed.
However, electrons can effectively transfer between different leads via second-order cotunneling processes.
These involve virtual transitions through the high-energy charged states of the island. 
The transition probability can be calculated using 2nd order perturbation theory \cite{dittrich1998quantum,fu2010Electron,qin2019transport}. 
Equivalently, below we derive the cotunneling-mediated effective Hamiltonian using a Schrieffer-Wolff transformation.
From central system Hamiltonian:
\begin{equation}
H_0 = H_C + H_T.
\label{eq:H_0_main to be schrieffer-wolff}
\end{equation}
By defining the fermion parity operator product as:
\begin{equation}
P_{0123} = i\gamma_0\gamma_1 i\gamma_2\gamma_3 .
\end{equation}

The charging Hamiltonian \eq{eq:H_C_strong cases} simplifies to:
\begin{equation}
H_C = E_C \frac{1 + P_{0123}}{2}.
\end{equation}
As established in \eq{eq:H_C_strong cases}, this charging energy creates distinct energy sectors: the low energy sector corresponds to states where $H_C = 0$ (i.e., $n_1 + n_2 = 1$), while the high energy sector corresponds to states where $H_C = E_C$ (i.e., $n_1 = n_2 = 0$ or $n_1 = n_2 = 1$). The tunneling Hamiltonian $H_T$ connects these energy sectors by flipping the fermion parity.
To derive the effective low-energy Hamiltonian, we apply a Schrieffer-Wolff transformation,
which transforms the Hamiltonian into a new basis where it is block-diagonal in these energy sectors perturbatively. 
 This transformation's generator is an anti-Hermitian operator $S$ that satisfies:
\begin{equation}
[S,H_C] = -H_T.
\label{eq:S_H_C}
\end{equation}

To find such operator $S$, we first note that $H_T$ alternates between the low and high energy sectors since it always flips the parity product, 
thus we can separate it into parts that raise and lower the energy:
\begin{equation}
H_T = H_T^+ + H_T^-,
\end{equation}
where $H_T^+$ raises the system from low to high energy sector, and $H_T^-$ lowers it from high to low energy sector.
Therefore, we can determine the form of $S$ as:
\begin{equation}
S = \frac{1}{E_C}(H_T^+ - H_T^-).
\label{eq:S_H_T + -}
\end{equation}
Then using the Baker-Campbell-Hausdorff formula, we have:
\begin{equation}
H' = e^S H e^{-S} = H + [S, H] + \frac{1}{2}[S,[S,H]] + \cdots.
\label{eq:BCH}
\end{equation}

Substituting our Hamiltonian components \eq{eq:H_0_main to be schrieffer-wolff} and \eq{eq:S_H_C} into \eq{eq:BCH}, we have:
\begin{equation}
H_0' = H_C   + \frac{1}{2}[S, H_T ] + \mathcal{O}(\frac{1}{E_C^2}H_T^3).
\label{eq:H_0_prime with schrieffer-wolff}
\end{equation}
Now, we substitute \eq{eq:S_H_T + -} into \eq{eq:H_0_prime with schrieffer-wolff}, we have:
\begin{equation}
H_0' = H_C + \frac{1}{E_C}[H_T^+, H_T^-] + \mathcal{O}(\frac{1}{E_C^2}H_T^3).
\label{eq:H_0_prime final with schrieffer-wolff}
\end{equation}
When acting \eq{eq:H_0_prime final with schrieffer-wolff} on the low-energy sector, using the fact that 
$H_C|\mathrm{low}\rangle = 0$ and $H_T^-|\mathrm{low}\rangle = 0$, we have:
\begin{equation}
H_{low} = -\frac{1}{E_C}H_T^- H_T^+ .
\label{eq:H_low_with schrieffer-wolff}
\end{equation}
These two operators $H_T^+$ and $H_T^-$ represent complementary virtual tunneling pathways:
\begin{itemize}
\item Pathway 1: Lead $\alpha' \rightarrow$ island ($H^+ = \lambda_{\alpha'}\gamma_{\alpha'}c_{\alpha',k}$) then island $\rightarrow$ Lead $\alpha$ ($H^- = \lambda_{\alpha}^{*}c_{\alpha,k}^{\dagger}\gamma_{\alpha}$)
\item Pathway 2: island $\rightarrow$ Lead $\alpha$ ($H^+ = \lambda_{\alpha}^{*}c_{\alpha,k}^{\dagger}\gamma_{\alpha}$) then Lead $\alpha' \rightarrow$ island ($H^- = \lambda_{\alpha'}\gamma_{\alpha'}c_{\alpha',k}$)
\end{itemize}
These pathways combine to yield the effective cotunneling Hamiltonian:

\begin{equation}
\tilde{H}_T =  \sum_{\alpha'=0}^{1} \sum_{\substack{\alpha=1 \\ \alpha > \alpha'}}^{2} \sum_{k, k'} (\mathcal{O}_{\alpha \alpha'} c_{\alpha k}^{\dagger} c_{\alpha' k'} + \text{H.c.}).
\label{eq:HtildeT_main}
\end{equation}
The MZM-dependent amplitude $\mathcal{O}_{\alpha \alpha'}$ is:
\begin{align} \label{eq:O_op_main}
\mathcal{O}_{\alpha \alpha'} &= t_{M,\alpha \alpha'} i \gamma_\alpha \gamma_{\alpha'},
\end{align}
where $t_{M,\alpha \alpha'} = i\frac{2 \lambda_{\alpha}^* \lambda_{\alpha'}}{E_c}$,
This operator describes the amplitude for an electron to effectively tunnel from lead $\alpha'$ to lead $\alpha$ via the MZM-assisted cotunneling pathway.

\section{Derivation of the Lindblad Master Equation} \label{app:deriv lindblad master equation}
To better understand the origin of the term \eq{eq:H_LS_main corresponding appendix} in the main text, 
we present a rigorous derivation of the Lindblad master equation 
for our system-environment interaction. We begin with the general form:
\begin{equation}
H_{I}=\sum_{\alpha\alpha'}A_{\alpha\alpha'}\otimes B_{\alpha\alpha'}.
\label{eq:H_I_tensor A B}
\end{equation}
Note that \eq{eq:HI_main total tunneling} can be cast into the above form
by identifying the system and bath operators as:
\begin{equation}
\begin{aligned}
A_{\alpha\alpha'} &= t_{R,\alpha \alpha'} + \mathcal{O}_{\alpha\alpha'} \\
B_{\alpha\alpha'} &= \sum_{k,k'}c_{\alpha,k}^\dagger c_{\alpha',k'}.
\end{aligned}
\label{eq:system_bath_operators}
\end{equation}
Applying the Born-Markov approximation,
we obtain the Lindblad master equation \cite{Breuer2002theory}:
\begin{equation}
\frac{d}{dt}\rho_{S}(t)=-i\left[H_{LS},\rho_{S}(t)\right]+\mathcal{D}\left(\rho_{S}(t)\right).
\label{eq:appendix_rho_master}
\end{equation}
The Lamb shift Hamiltonian $H_{LS}$ and dissipator $\mathcal{D}(\rho_{S})$ are given by:
\begin{equation}
H_{LS}=\sum_{\omega}\sum_{\alpha\alpha^{\prime}}S_{\alpha\alpha^{\prime}}(\omega)A_{\alpha\alpha^{\prime}}(-\omega)A_{\alpha^{\prime}\alpha}(\omega),
\label{eq:H_LS_main}
\end{equation}
\begin{equation}
\begin{aligned}
\mathcal{D}\left(\rho_{S}\right) = \sum_{\omega}\sum_{\alpha\alpha^{\prime}}&\Gamma_{\alpha\alpha^{\prime}}(\omega) \Big(A_{\alpha^{\prime}\alpha}(\omega)\rho_{S}A_{\alpha\alpha^{\prime}}(-\omega) \\
&-\frac{1}{2}\left\{ A_{\alpha\alpha^{\prime}}(-\omega)A_{\alpha^{\prime}\alpha}(\omega),\rho_{S}\right\} \Big),
\end{aligned}
\label{eq:D_rho_main}
\end{equation}
where $\{,\}$ denotes the anti-commutator.
$S_{\alpha\alpha^{\prime}}(\omega)$ and $\Gamma_{\alpha\alpha^{\prime}}(\omega)$ 
, which determine the strength of the coherent and dissipative dynamics, are calculated from the bath correlation functions:
\begin{equation}
\Gamma_{\alpha\alpha^{\prime}}(\omega) = 2\text{Re}(\Gamma_{\alpha\alpha^{\prime}\alpha^{\prime}\alpha}(\omega)),
\end{equation}
\begin{equation}
S_{\alpha\alpha^{\prime}}(\omega) = \text{Im}(\Gamma_{\alpha\alpha^{\prime}\alpha^{\prime}\alpha}(\omega)),
\end{equation}
where $\Gamma_{\alpha\alpha'\beta\beta'}(\omega)$ is the bath correlation function defined as:
\begin{equation}
\Gamma_{\alpha\alpha'\beta\beta'}(\omega) = \int_{0}^{\infty} e^{i\omega t} \langle B_{\alpha\alpha'}(t)B_{\beta\beta'}(0)\rangle dt .
\end{equation}
As a demonstration, in the following we set all bias
$V_\alpha=0$ and calculate $S_{\alpha\alpha'},\Gamma_{\alpha\alpha'}$ near zero temperature, introduce $\delta\xi = \xi-\xi^{\prime}$:
\begin{equation}
\begin{aligned}
S_{12} &= \nu^{2}\underset{|\xi|,|\xi^{\prime}|\leq\Lambda/4}{\iint}d\xi d\xi^{\prime}\int_{0}^{\infty}dt\sin(\delta\xi\cdot t) \\
&\times n(\xi^{\prime}+\delta\xi)\left(1-n(\xi^{\prime})\right) \\
&= \nu^{2}\mathbb{P}\int_{-\frac{\Lambda}{2}}^{\frac{\Lambda}{2}}d\delta\xi\frac{1}{\delta\xi}\frac{\delta\xi}{\exp(\frac{\delta\xi}{T})-1} \\
&\overset{T\rightarrow0}{\approx} -\frac{\Lambda \nu^2}{2}.
\end{aligned}
\end{equation}
Note that in the above derivation, we have used the following relation derived from improper integration:
\begin{equation}
\int_{0}^{\Omega}\int_{0}^{\infty}\sin((\omega-\omega_j)t)dt F(\omega)d\omega = \mathbb{P}\int_{0}^{\Omega}\frac{F(\omega)}{\omega-\omega_j}d\omega .
\end{equation}

Similar calculation can be done for $S_{01}$ and $S_{10}$:
\begin{equation}
\begin{aligned}
S_{12} = S_{21} &\approx -\frac{\Lambda \nu^2}{2}, \\
S_{01} = S_{02} &\approx -\frac{\Lambda \nu^2}{2},  \\
S_{10} = S_{20} &\approx -\frac{\Lambda \nu^2}{2}, 
\end{aligned}
\label{eq:S_01_02_10_20}
\end{equation}
where $\Lambda$ is the bandwidth cutoff, $\nu$ is the density of states of the leads.
Under the same condition, we have:
\begin{equation}
\begin{aligned}
\Gamma_{21} = \Gamma_{12} &= 2\pi\iint d\xi d\xi^{\prime}\nu(\xi)\nu(\xi^{\prime}) \\
&\times n(\xi)\left(1-n(\xi^{\prime})\right)\delta(\xi-\xi^{\prime})\\
&= \pi \nu^{2}\int d\delta\xi\frac{\delta\xi}{\exp(\frac{\delta\xi}{T})-1}\delta(\delta\xi)\\
&= 2\pi \nu^{2}\lim_{\delta\xi\rightarrow0}\frac{\delta\xi}{\exp(\frac{\delta\xi}{T})-1} = 2\pi \nu^{2}T .
\end{aligned}
\end{equation}

Similarly, we have:
\begin{equation}
\begin{aligned}
\Gamma_{21} = \Gamma_{12} &= 2\pi \nu^{2}T, \\
\Gamma_{01} = \Gamma_{02} &= 2\pi \nu^{2}T, \\
\Gamma_{10} = \Gamma_{20} &= 2\pi \nu^{2}T .
\end{aligned}
\label{eq:gamma_01_02_10_20}
\end{equation}

Using these calculated rates and energy shifts, we obtain the specific form of master equation as follows:

The Lamb shift Hamiltonian, which gives rise to the effective MZM coupling, can be calculated by substituting \eq{eq:S_01_02_10_20} and \eq{eq:system_bath_operators} into \eq{eq:H_LS_main}:
\begin{equation}
H_{LS} \approx-\nu^2\Lambda\sum_{\alpha'=0}^{1} \sum_{\substack{\alpha=1 \\ \alpha > \alpha'}}^{2}\left(t_{R,\alpha \alpha'}^*\mathcal{O}_{\alpha\alpha^{\prime}} + \text{H.c.}\right).
\label{eq:H_LS_appendix_calculation}
\end{equation}
By substituting \eq{eq:gamma_01_02_10_20} into \eq{eq:D_rho_main}, the dissipator $\mathcal{D}(\rho_S)$ in the master equation simplifies to:
\begin{equation}
\begin{aligned}
\mathcal{D}(\rho_S) &= \sum_{\substack{\alpha\neq\alpha'}} 2\pi \nu^2 T \Big( A_{\alpha'\alpha} \rho_S A_{\alpha\alpha'} \\
&- \frac{1}{2} \{ A_{\alpha\alpha'} A_{\alpha'\alpha}, \rho_S \} \Big).
\end{aligned}
\label{eq:D_rho_appendix_calculation}
\end{equation}
To derive the matrix representation of the master equation, we first establish the operator basis for our physical system. For a 3-Majorana zero mode (MZM) system ($\gamma_0, \gamma_1, \gamma_2$), the parity operators form a Clifford algebra that admits a faithful representation through Pauli matrices:
\begin{equation}
\sigma_1 = i\gamma_1\gamma_0,\quad \sigma_2 = i\gamma_2\gamma_0,\quad \sigma_3 = i\gamma_2\gamma_1.
\label{eq:sigma_1_2_3 of mzms}
\end{equation}
The density matrix can be decomposed in this Pauli basis as:
\begin{equation}
\rho_S(t) = \frac{1}{2}\left(\mathbb{I} + \sum_{\mu=1}^{3} \rho_{\mu}(t) \sigma_{\mu}\right),
\end{equation}
where the Bloch vector components $\rho_{\mu}(t) = \text{Tr}(\rho_S(t) \sigma_{\mu})$ completely describe the quantum state. 
This single-qubit parametrization remains valid and complete because we work in the low-energy sector 
where the total parity product $i\gamma_0\gamma_1i\gamma_2\gamma_3$ is conserved, reducing the 
Hilbert space to two dimensions; This can also be seen from the fact that the Lindblad 
operators in \eq{eq:D_rho_appendix_calculation} and coherent evolution terms 
in \eq{eq:H_LS_appendix_calculation} generate dynamics entirely within the Pauli
 operator space defined above. 
 
Substituting this parametrization into \eq{eq:D_rho_appendix_calculation} and \eq{eq:H_LS_appendix_calculation} transforms it into a system of linear ordinary differential equations (ODEs) for the Bloch vector components:
\begin{equation}
\frac{d}{dt} \bm{\rho}(t) = M(t) \bm{\rho}(t),
\label{eq:bloch_vector_ode}
\end{equation}
where the matrix $M$ is given by:
\begin{equation}
M = \left(\begin{array}{cccc}
0 & 0 & 0 & 0 \\
0 & -\varGamma_{20}-\varGamma_{21} & -h_{3} & h_{2} \\
0 & h_{3} & -\varGamma_{10}-\varGamma_{21} & -h_{1} \\
0 & -h_{2} & h_{1} & -\varGamma_{10}-\varGamma_{20}
\end{array}\right).
\label{eq:M_matrix_structure}
\end{equation}

In our numerical calculation, without loss of generality, we set $t_{M,\alpha \alpha'}$ as a real number. 
Then the dissipation rates $\varGamma_{\alpha\beta}$ and coherent coupling terms $h_{\alpha}$ are given by:
\begin{align}
\varGamma_{\alpha \alpha'} &= 4 \pi \nu^2 T t_{M,\alpha \alpha'}^2, \\
h_{\alpha}(t) &\approx -4 \nu^2 \Lambda \operatorname{Re}\left( t_{R,\alpha\beta}^{*} t_{M,\alpha\beta} \right),
\end{align}
where $(\alpha,\beta) \in \{(1,0), (2,0), (2,1)\}$ for $h_1$, $h_2$, and $h_3$ respectively.

\section{Majorana Zero Mode Braiding} \label{app:mzm_braiding}
In this appendix, we provide a detailed description of the braiding processes for 
Majorana zero modes (MZMs) in different junction configurations and derive the braiding operators. 
Fig.~\ref{fig:braiding_comparison} illustrates two different approaches to MZM braiding. 
The branch cuts represent the phase discontinuities in the wavefunction description of MZMs. 
When one MZM crosses another's branch cut, it picks up a minus sign in its transformation, 
reflecting the fundamental non-Abelian statistics of MZMs.

To understand the mathematical foundation of the non-Abelian braiding effect and prove 
the results in Fig.~\ref{fig:braiding_comparison}, 
we model the Y junction in Fig.~\ref{fig:braiding_yjunction} by the Hamiltonian \cite{alicea2011non}:
\begin{equation}
H = i \sum_{j=1}^3 \Delta_{0j} \gamma_0 \gamma_j.
\label{eq:H_Y_junction}
\end{equation}

The parity operators satisfying Clifford algebra can be represented by Pauli matrices:
\begin{align}
\tau_z &= i\gamma_0\gamma_3, \quad \tau_y = i\gamma_3\gamma_2, \quad \tau_x = i\gamma_2\gamma_0.
\end{align}
The total parity operator is defined as:
\begin{equation}
\pi_z = -\gamma_0\gamma_1\gamma_2\gamma_3.
\end{equation}

We can express the Hamiltonian \eq{eq:H_Y_junction} as:
\begin{equation}
H = \Delta_{03}\tau_z - \Delta_{01}\tau_y\pi_z - \Delta_{02}\tau_x.
\label{eq:Hamiltonian for berry phase}
\end{equation}

The braiding of Y junction configuration in Fig.~\ref{fig:braiding_yjunction} is achieved by adiabatically evolving the system through the sequence:
\begin{equation}
\Delta_{03} \rightarrow \Delta_{01} \rightarrow \Delta_{02} \rightarrow \Delta_{03}.
\label{eq:Y_braiding_sequence}
\end{equation}

On the Bloch sphere, this corresponds to the $\vec{d}$ vector winding around a solid angle of magnitude $\pi/4$. 
The winding direction depends on the initial state and the sign of the $\tau_y$ coefficient in Hamiltonian \eq{eq:Hamiltonian for berry phase}. 
For sequence \eq{eq:Y_braiding_sequence}, the initial state is an eigenstate of $\tau_z$, and for initial states $\tau_z = +1$ or $\tau_z = -1$, the winding directions are opposite. Similarly, when the sign of the $\tau_y$ coefficient in the Hamiltonian changes, the winding direction also reverses.
The resulting unitary transformation is:
\begin{align}
U_{12}=B_{21} &= e^{-\pi_z\tau_z\frac{i\pi}{4}} \\
&= e^{\gamma_0\gamma_1\gamma_2\gamma_3\gamma_0\gamma_3\frac{\pi}{4}} \\
&= e^{\frac{\pi\gamma_2\gamma_1}{4}} \\
&= \frac{1 + \gamma_2\gamma_1}{\sqrt{2}}. 
\end{align}
It can be verified that this transformation exchanges the Majorana operators as:
\begin{align}
B_{21}\gamma_1B_{21}^\dagger &= \gamma_2, \\
B_{21}\gamma_2B_{21}^\dagger &= -\gamma_1.
\end{align}
For the triangle junction in Fig.~\ref{fig:braiding_trijunction}, the Hamiltonian is given by:
\begin{equation}
H = \sum_{\alpha'=0}^{1} \sum_{\substack{\alpha=1 \\ \alpha > \alpha'}}^{2} \sum_{k, k'}(t_{R,\alpha \alpha'} + \mathcal{O}_{\alpha\alpha'})(c_{\alpha,k}^\dagger c_{\alpha',k'} + \text{H.c.}).
\label{eq:H_triangle_junction}
\end{equation}

Repeat \eq{eq:sigma_1_2_3 of mzms} here:
\begin{equation}
  \sigma_x = i\gamma_1\gamma_0,\quad \sigma_y = i\gamma_2\gamma_0,\quad \sigma_z = i\gamma_2\gamma_1.
  \end{equation}

We can express the Hamiltonian \eq{eq:H_triangle_junction} as:
\begin{equation}
H = \Delta_{10}\sigma_x + \Delta_{20}\sigma_y + \Delta_{21}\sigma_z.
\end{equation}

The braiding of triangle junction configuration in Fig.~\ref{fig:braiding_trijunction} is achieved by adiabatically evolving the system through the sequence:
\begin{equation}
\Delta_{21} \rightarrow \Delta_{10} \rightarrow \Delta_{20} \rightarrow \Delta_{21}.
\label{eq:triangle_braiding_sequence}
\end{equation}

On the Bloch sphere, this corresponds to the $\vec{d}$ vector winding around a solid angle of magnitude $\pi/4$. 
The winding direction depends on the initial state.
For sequence \eq{eq:triangle_braiding_sequence}, the initial state is an eigenstate of $\sigma_z$, and for initial states $\sigma_z = +1$ or $\sigma_z = -1$, the winding directions are opposite.
The resulting unitary transformation is:
\begin{align}
U_{12}=B_{12} &= e^{\sigma_z\frac{i\pi}{4}} \\
&= e^{-\frac{\pi\gamma_2\gamma_1}{4}} \\
&= \frac{1 + \gamma_1\gamma_2}{\sqrt{2}} .
\end{align}
It can be verified that this transformation exchanges the Majorana operators as:
\begin{align}
B_{12}\gamma_1B_{12}^\dagger &= - \gamma_2, \\
B_{12}\gamma_2B_{12}^\dagger &= \gamma_1,
\end{align}
which matches the result in Fig.~\ref{fig:braiding_comparison}.

As stated in Sec.~\ref{subsec:interference_coupling}, 
the correspondence between the coherent coupling strength $\Delta_{\alpha\alpha'}$ and 
quantum interference contribution to conductance $\Delta G_{\alpha\alpha'}$ in the interference experiment provides us with the feasibility of 
fine-tuning the coherent coupling strength. 
This enables us to implement the following braiding sequence:
\begin{equation}
\Delta_{21} \rightarrow \Delta_{10} \rightarrow (\Delta_{10}=\Delta_\perp\cos\theta,\ \Delta_{20}=\Delta_\perp\sin\theta) \rightarrow \Delta_{21},
\label{eq:theta_braiding_sequence}
\end{equation}
where the 3rd step means that $\Delta_{10}$ and $\Delta_{20}$ are both non-zero, and their ratio is given by:
\begin{equation}
\frac{\Delta_{20}}{\Delta_{10}} = \tan(\theta).
\end{equation}

\begin{figure}[htp!] 
  \includegraphics[width=0.6\textwidth]{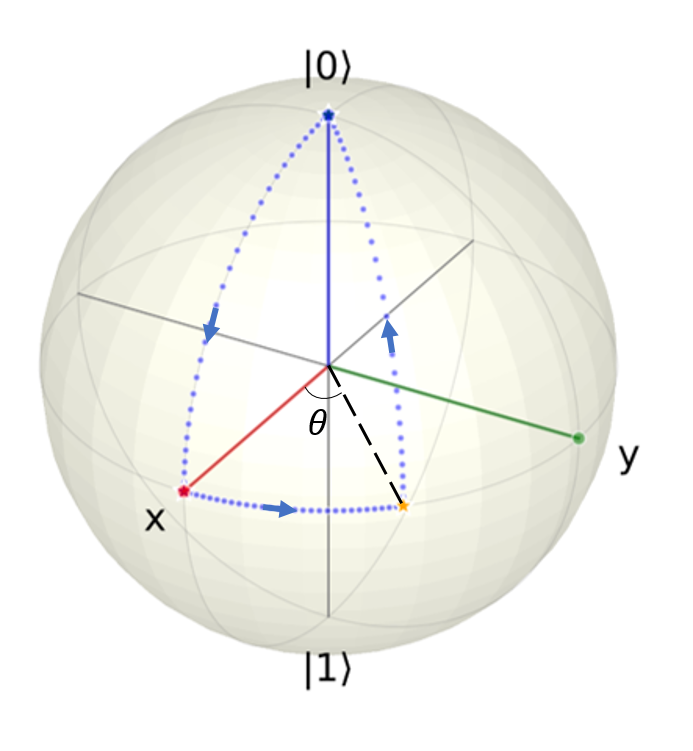}
  \caption {Visualization of the sequence \eq{eq:theta_braiding_sequence} on the Bloch sphere, showing the path of the $\vec{d}$ vector during the operation.
  Specifically, $\theta = \pi/2$ corresponds to $B_{12}(\pi/2)$; 
  $\theta = \pi/4$ to $B_{12}(\pi/4)$, employed in echoed Clifford gates; 
  and $\theta = \pi/8$ to $B_{12}(\pi/8)$, employed in the echoed $T$ gate.
 }
 \label{fig:braiding_sequence_bloch}
\end{figure}

\begin{figure*}[htp!]
  \centering
  \subfloat[]{\begin{overpic}[width=0.47\textwidth]{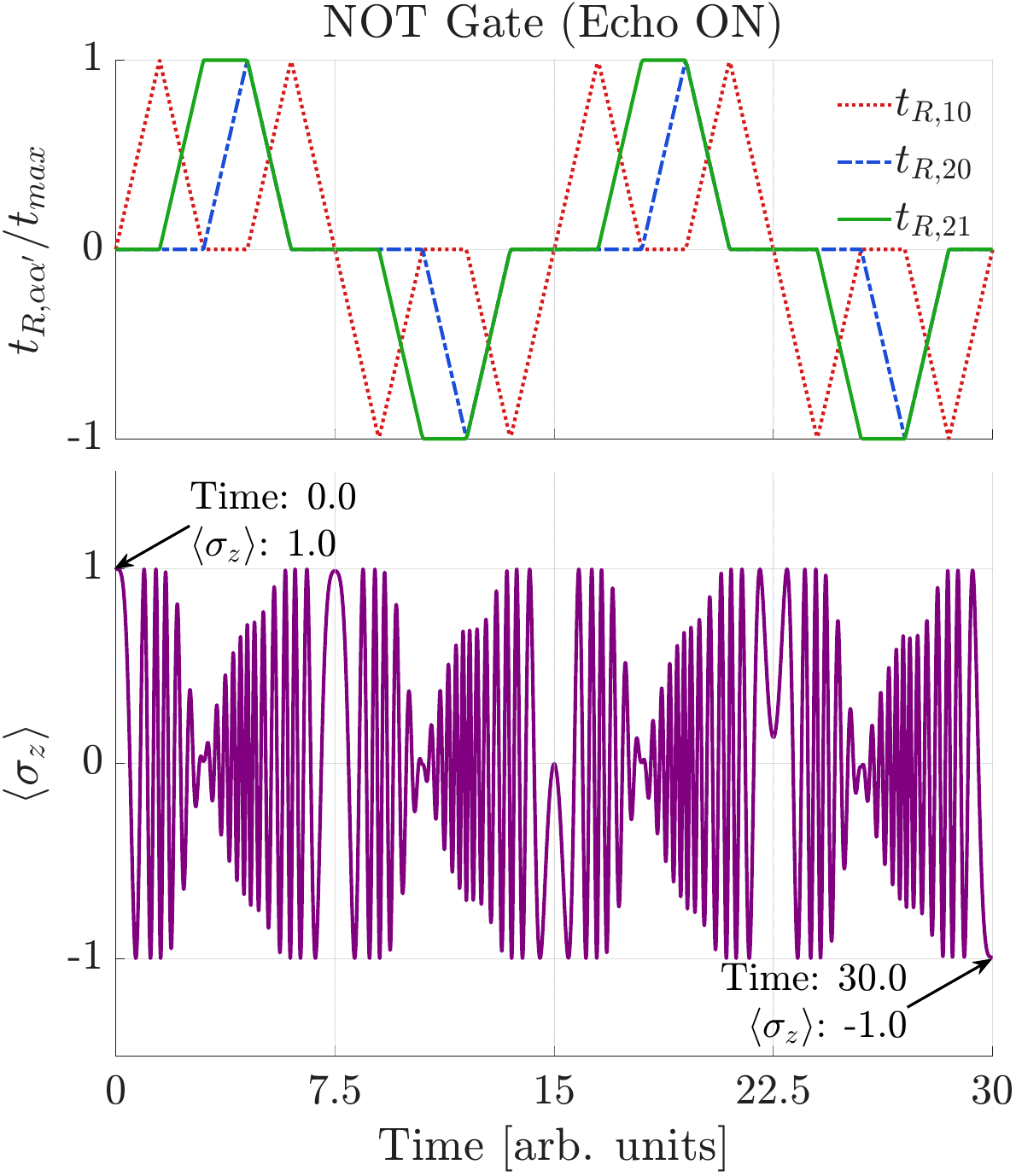}\put(0,97){\textbf{(a)}}\end{overpic}\label{fig:modulation_linear}}
  \hspace{0.03\textwidth}
  \subfloat[]{\begin{overpic}[width=0.47\textwidth]{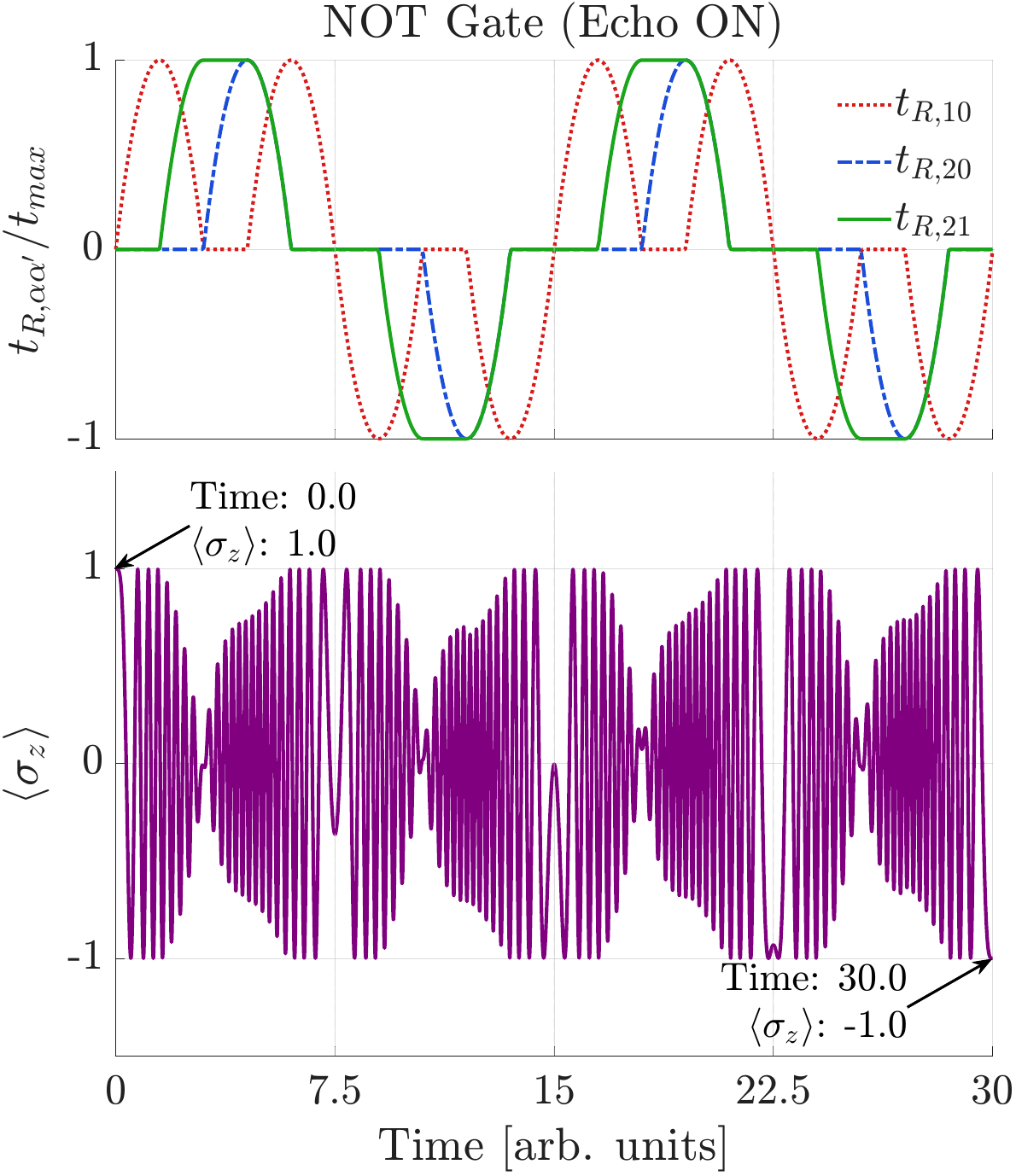}\put(0,97){\textbf{(b)}}\end{overpic}\label{fig:modulation_quadratic}}\\
  \subfloat[]{\begin{overpic}[width=0.47\textwidth]{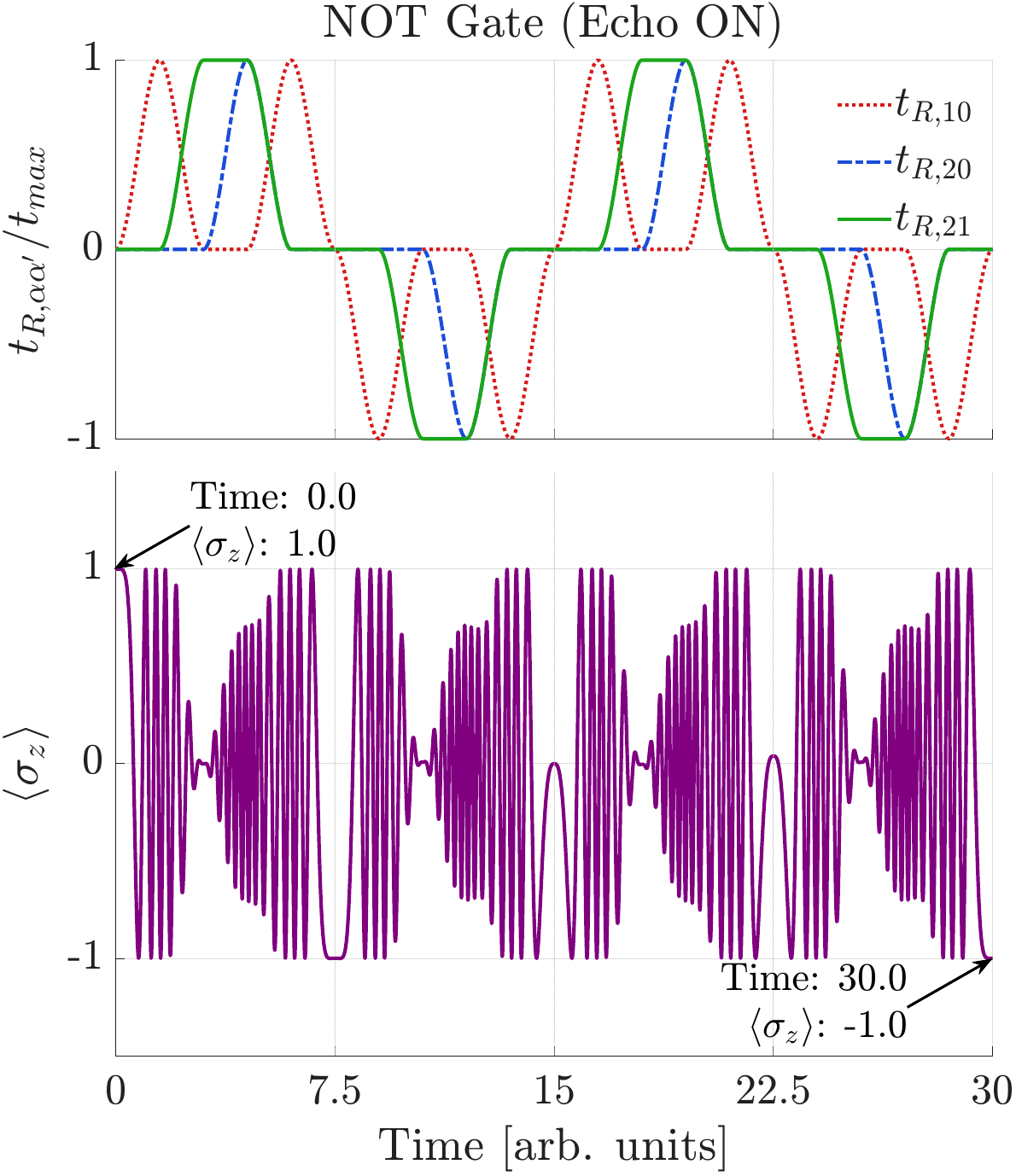}\put(0,97){\textbf{(c)}}\end{overpic}\label{fig:modulation_cubic}}
  \hspace{0.03\textwidth}
  \subfloat[]{\begin{overpic}[width=0.47\textwidth]{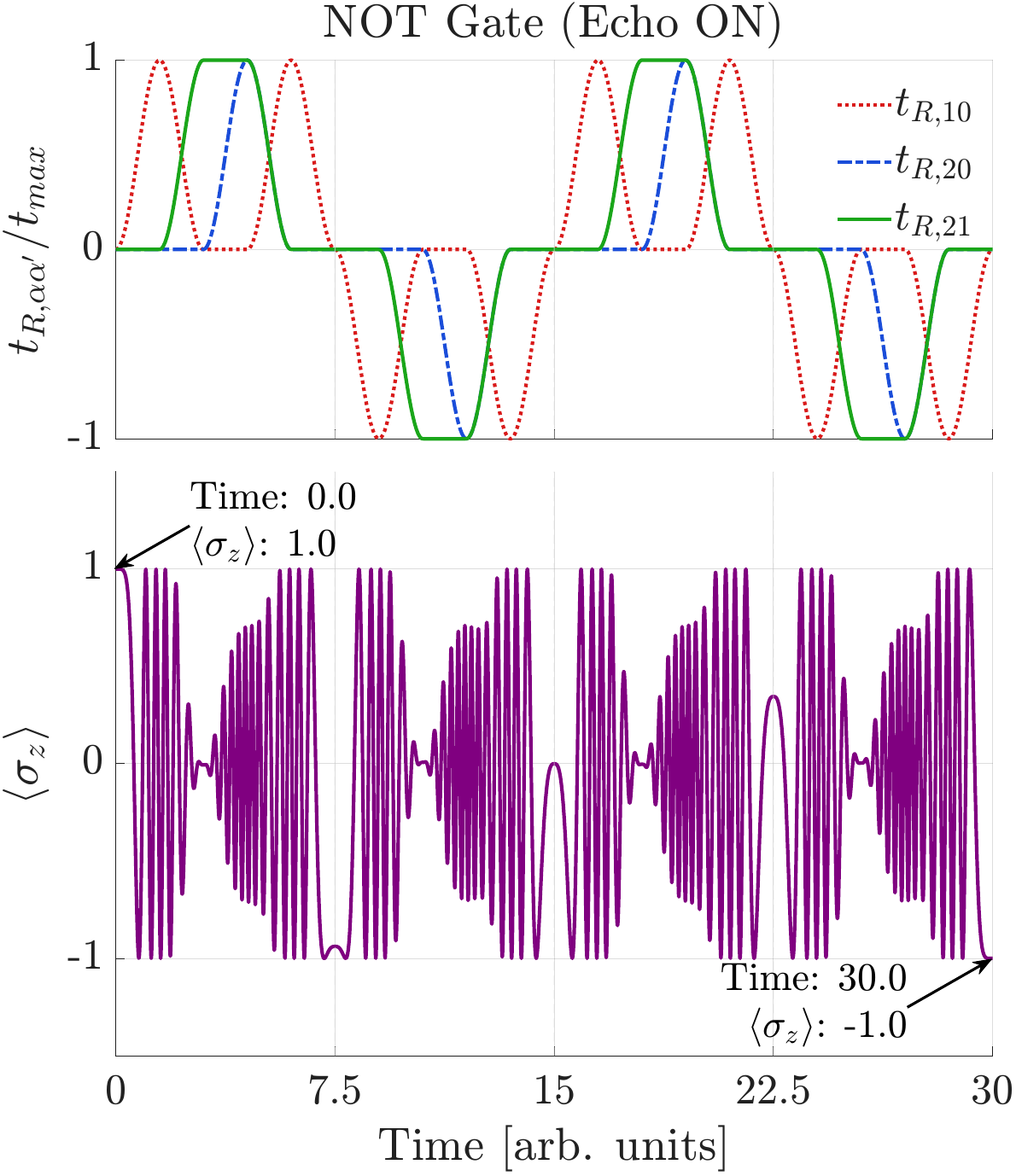}\put(0,97){\textbf{(d)}}\end{overpic}\label{fig:modulation_cosine}}
  \caption{Simulation results of the NOT gate operation using different $t_{R,\alpha \alpha'}$ modulation functions, simulation parameters are identical to those used in Fig.~\ref{fig:not_gate_simulation_long}. The subfigures: (a) Linear $t_{R,\alpha \alpha'}$ modulation function, (b) quadratic $t_{R,\alpha \alpha'}$ modulation function, (c) cubic $t_{R,\alpha \alpha'}$ modulation function, and (d) cosine $t_{R,\alpha \alpha'}$ modulation function. As depicted in the upper panel of each subfigure, $t = 0$ to $7.5$ corresponds to $B_{10}(\pi/4)$, $t = 7.5$ to $15$ represents the echoed $B_{10}(\pi/4)$ which combines with the first $B_{10}(\pi/4)$ to form the complete first $B_{10}(\pi/2)$ operation, and $t = 15$ to $30$ implements the second $B_{10}(\pi/2)$ operation similarly. Within the first segment ($t = 0$ to $7.5$): (i) $t = 0$ to $1.5$, $t_{R,10}$ ramps up while other parameters remain zero; (ii) $t = 1.5$ to $3$, $t_{R,10}$ decreases to zero, $t_{R,21}$ increases, and $t_{R,20}$ remains zero; (iii) $t = 3$ to $4.5$, $t_{R,10}$ remains zero, $t_{R,21}$ maintains its maximum value, $t_{R,20}$ increases to the same maximum value; (iv) $t = 4.5$ to $6$, both $t_{R,20}$ and $t_{R,21}$ decrease while $t_{R,10}$ increases; (v) $t = 6$ to $7.5$, $t_{R,10}$ decreases while $t_{R,20}$ and $t_{R,21}$ remain zero.}
  \label{fig:different_modulations}
\end{figure*}

\begin{figure*}[htp!]
  \centering
  \subfloat[]{\begin{overpic}[width=0.47\textwidth]{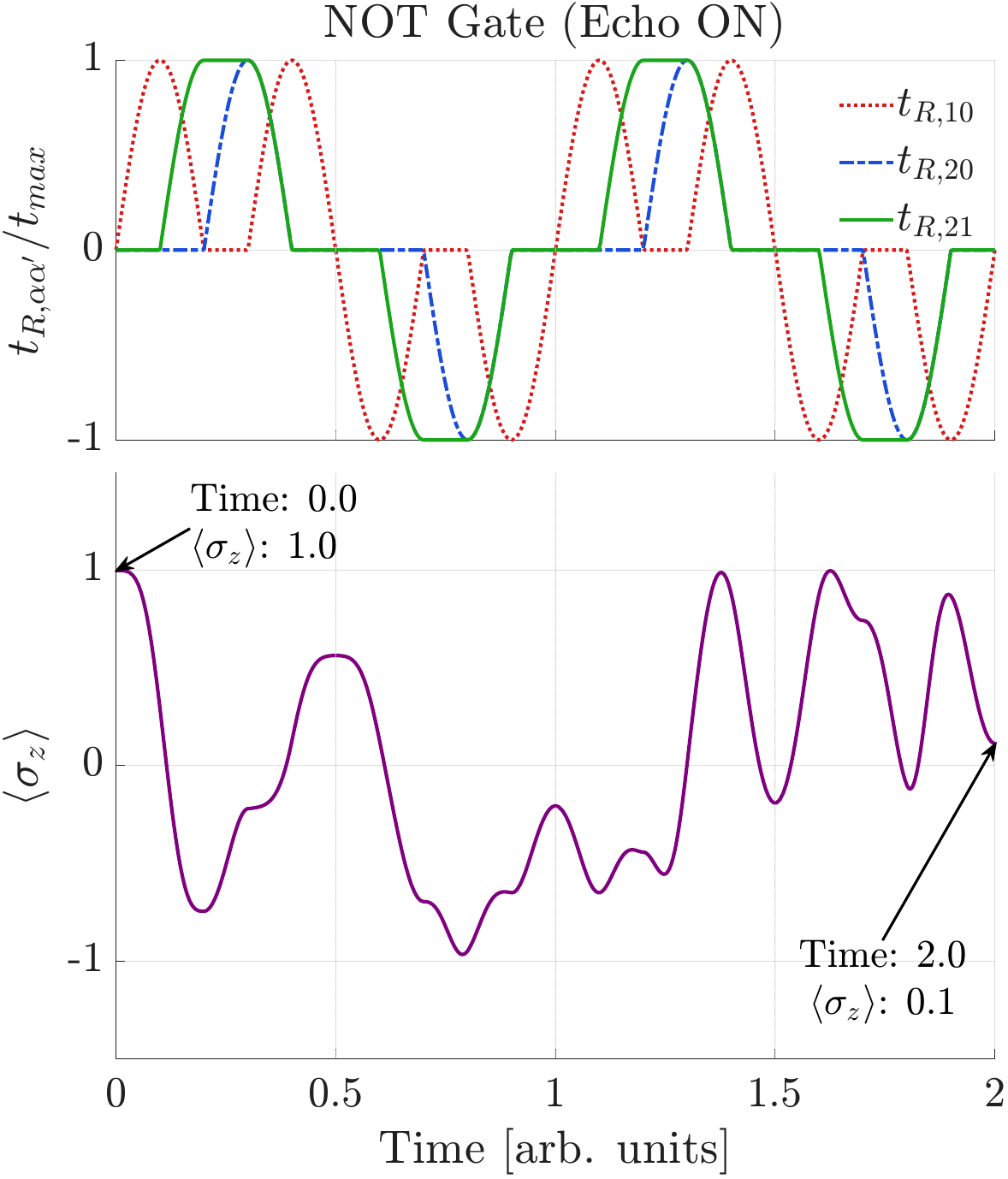}\put(0,97){\textbf{(a)}}\end{overpic}\label{fig:T_step_0p1}}
  \hspace{0.03\textwidth}
  \subfloat[]{\begin{overpic}[width=0.47\textwidth]{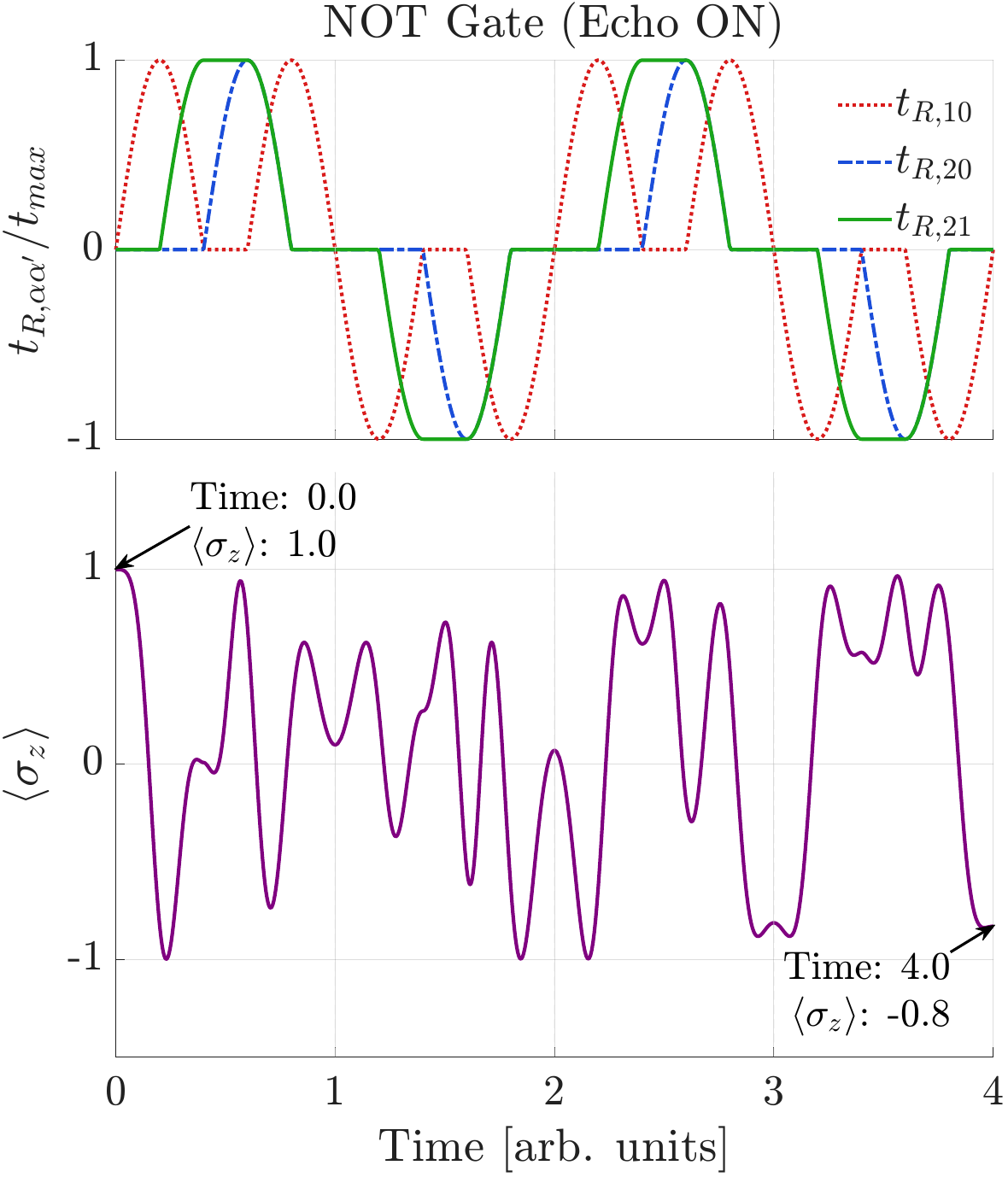}\put(0,97){\textbf{(b)}}\end{overpic}\label{fig:T_step_0p2}}\\
  \subfloat[]{\begin{overpic}[width=0.47\textwidth]{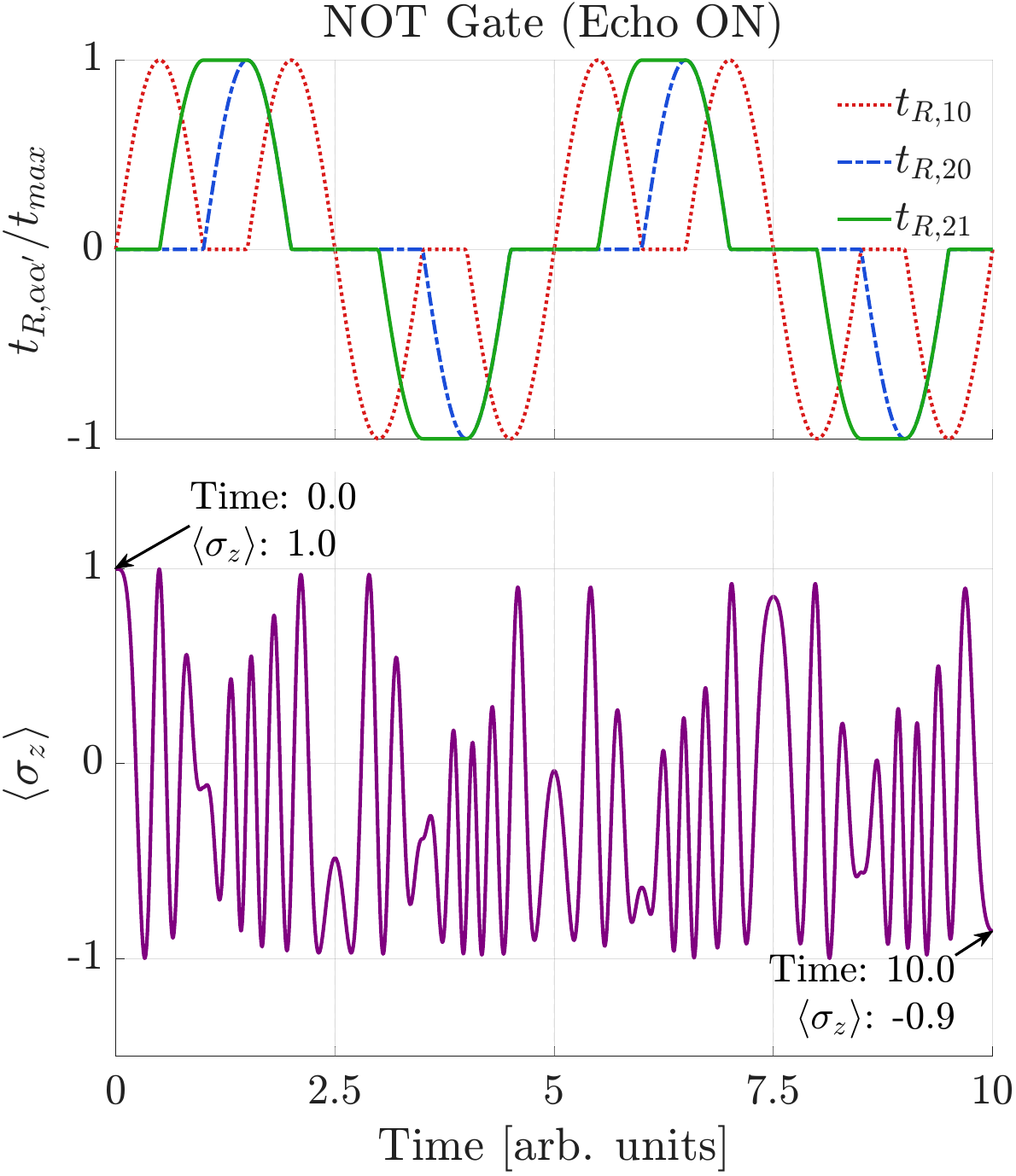}\put(0,97){\textbf{(c)}}\end{overpic}\label{fig:T_step_0p5}}
  \hspace{0.03\textwidth}
  \subfloat[]{\begin{overpic}[width=0.47\textwidth]{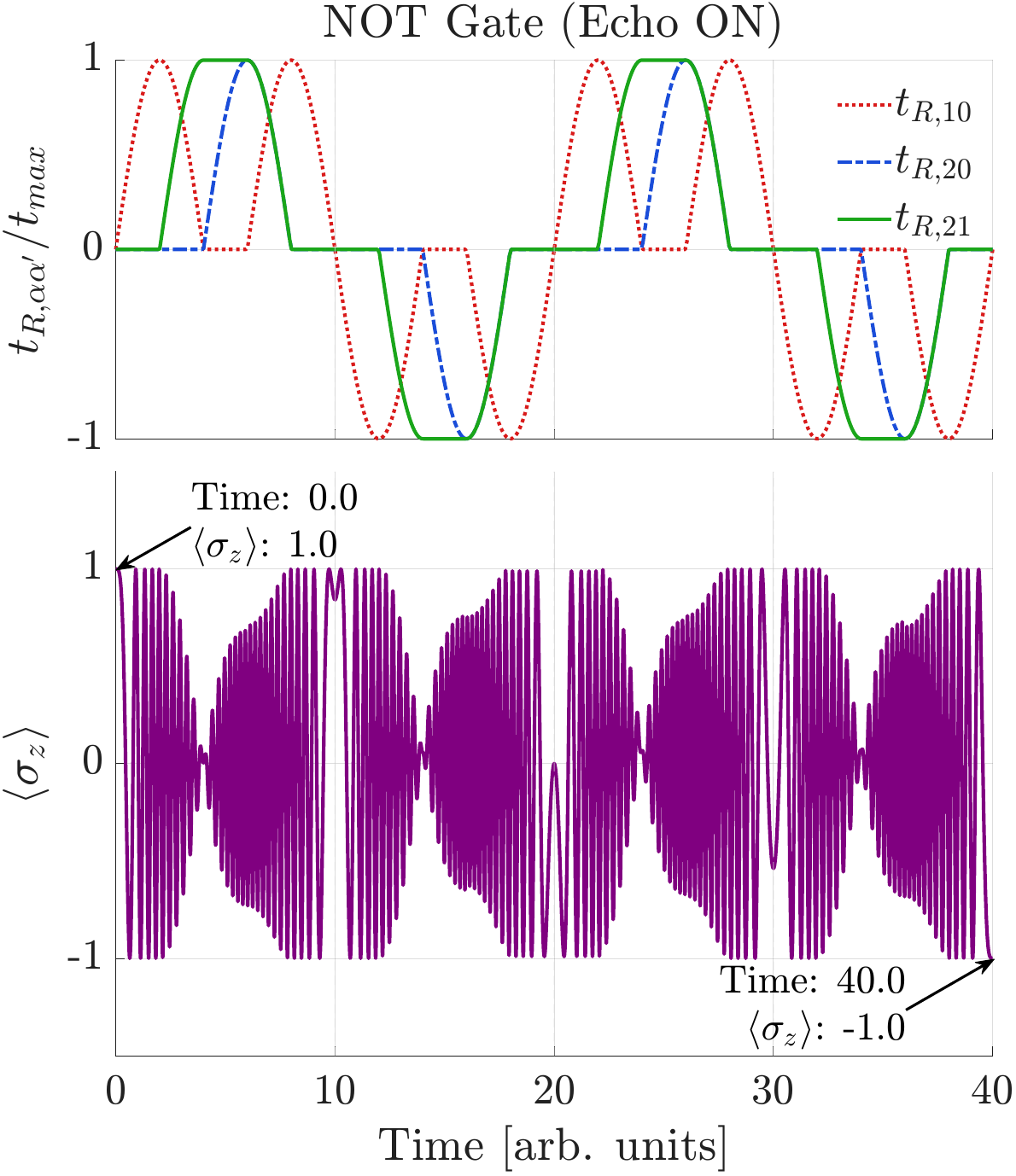}\put(0,97){\textbf{(d)}}\end{overpic}\label{fig:T_step_10p0}}
  \caption{Simulation results of the NOT gate operation with different segment operation times $T_{\mathrm{half}}$, the other simulation parameters are identical to those used in Fig.~\ref{fig:not_gate_simulation_long}. The subfigures: (a) $T_{\mathrm{half}} = 0.5$ (non-adiabatic), (b) $T_{\mathrm{half}} = 1.0$ (partially non-adiabatic), (c) $T_{\mathrm{half}} = 2.5$ (adiabatic), and (d) $T_{\mathrm{half}} = 10.0$ (fully adiabatic). For very short operation times ($T_{\mathrm{half}} = 0.5$), the adiabaticity condition is violated, resulting in significant deviations from the ideal NOT gate behavior. At $T_{\mathrm{half}} = 1.0$, some improvement is observed but the operation remains partially non-adiabatic. By $T_{\mathrm{half}} = 2.5$, the system more closely follows the expected trajectory, and at $T_{\mathrm{half}} = 10.0$, the operation achieves high fidelity.}
  \label{fig:different_T_steps}
  \end{figure*}

\begin{figure*}[htp!]
  \centering
  \subfloat[]{\begin{overpic}[width=0.47\textwidth]{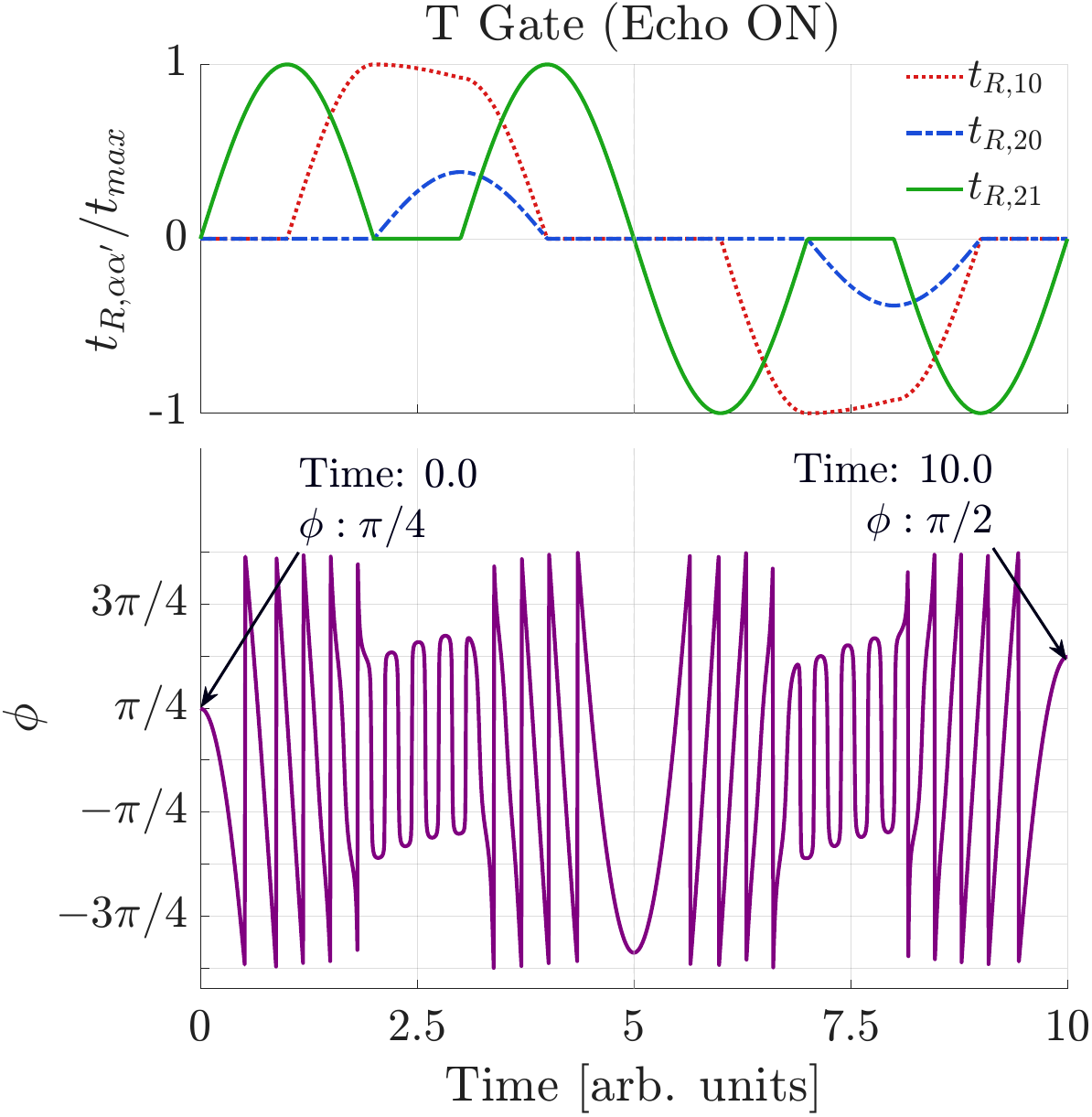}\put(0,97){\textbf{(a)}}\end{overpic}\label{fig:tgate_initial_state2_combined}}
  \hspace{0.03\textwidth}
  \subfloat[]{\begin{overpic}[width=0.47\textwidth]{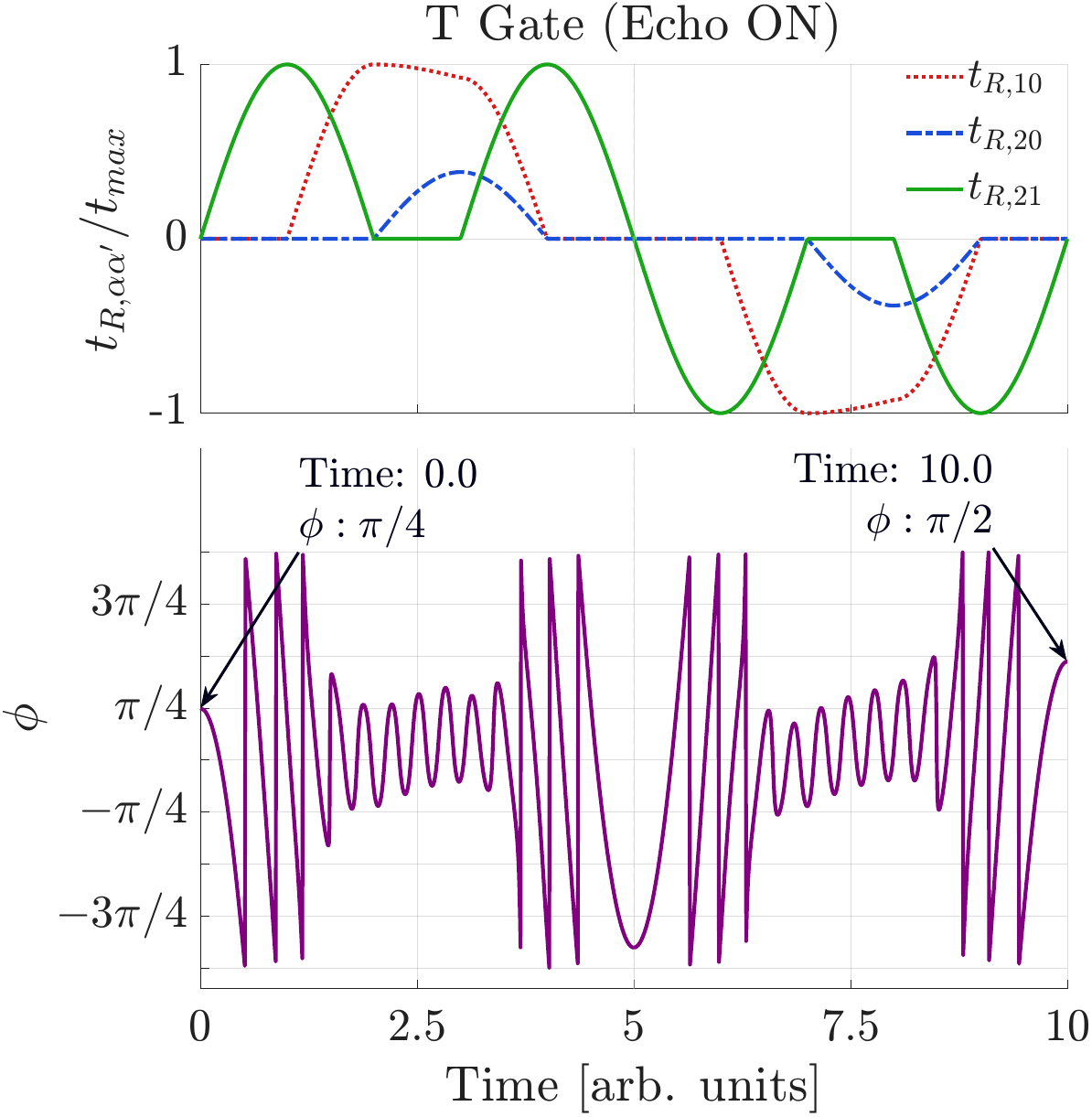}\put(0,97){\textbf{(b)}}\end{overpic}\label{fig:tgate_initial_state1_combined}}\\
  \subfloat[]{\begin{overpic}[width=0.47\textwidth,trim=100 140 100 170,clip]{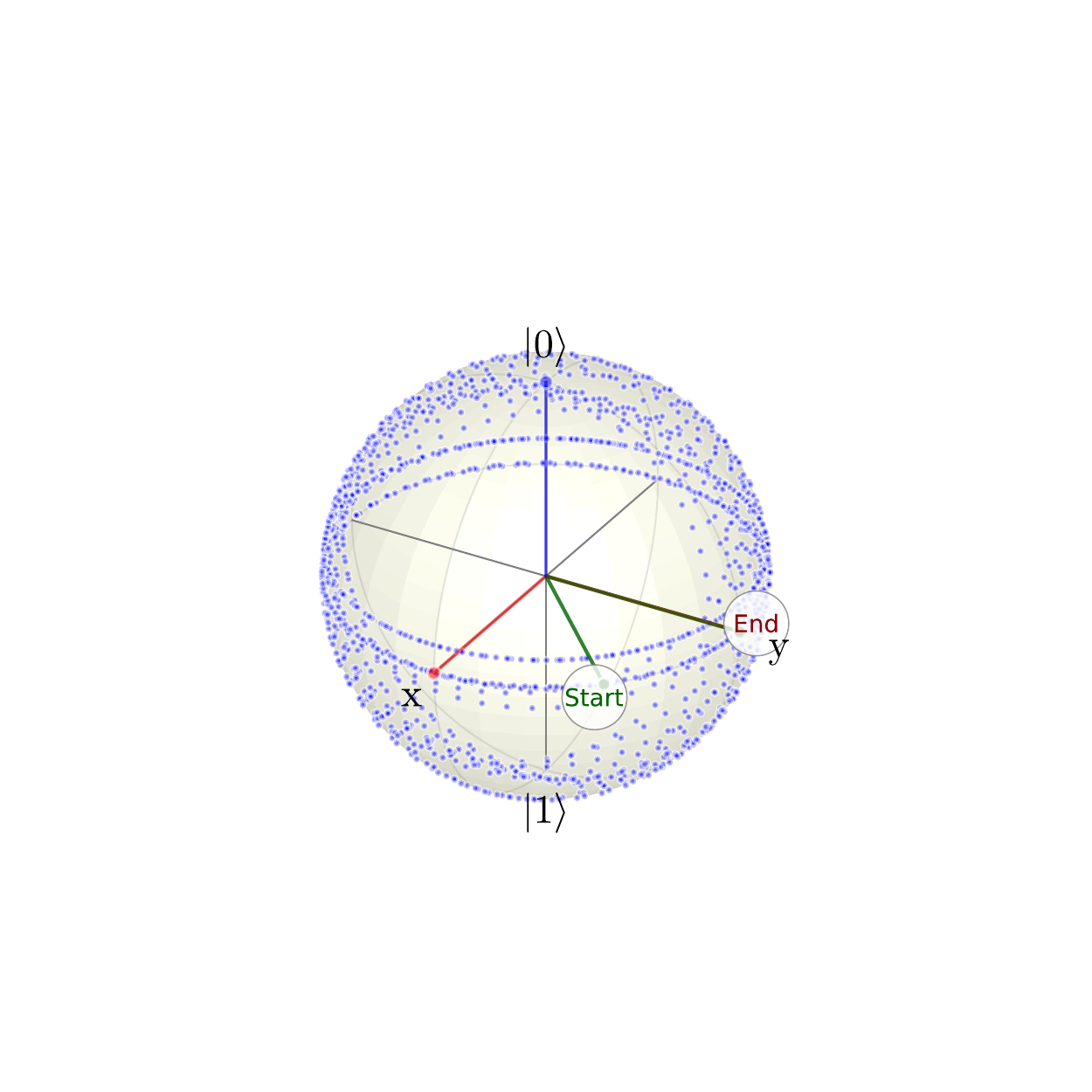}\put(0,72){\textbf{(c)}}\end{overpic}\label{fig:tgate_initial_state2_bloch}}
  \hspace{0.03\textwidth}
  \subfloat[]{\begin{overpic}[width=0.47\textwidth,trim=100 140 100 170,clip]{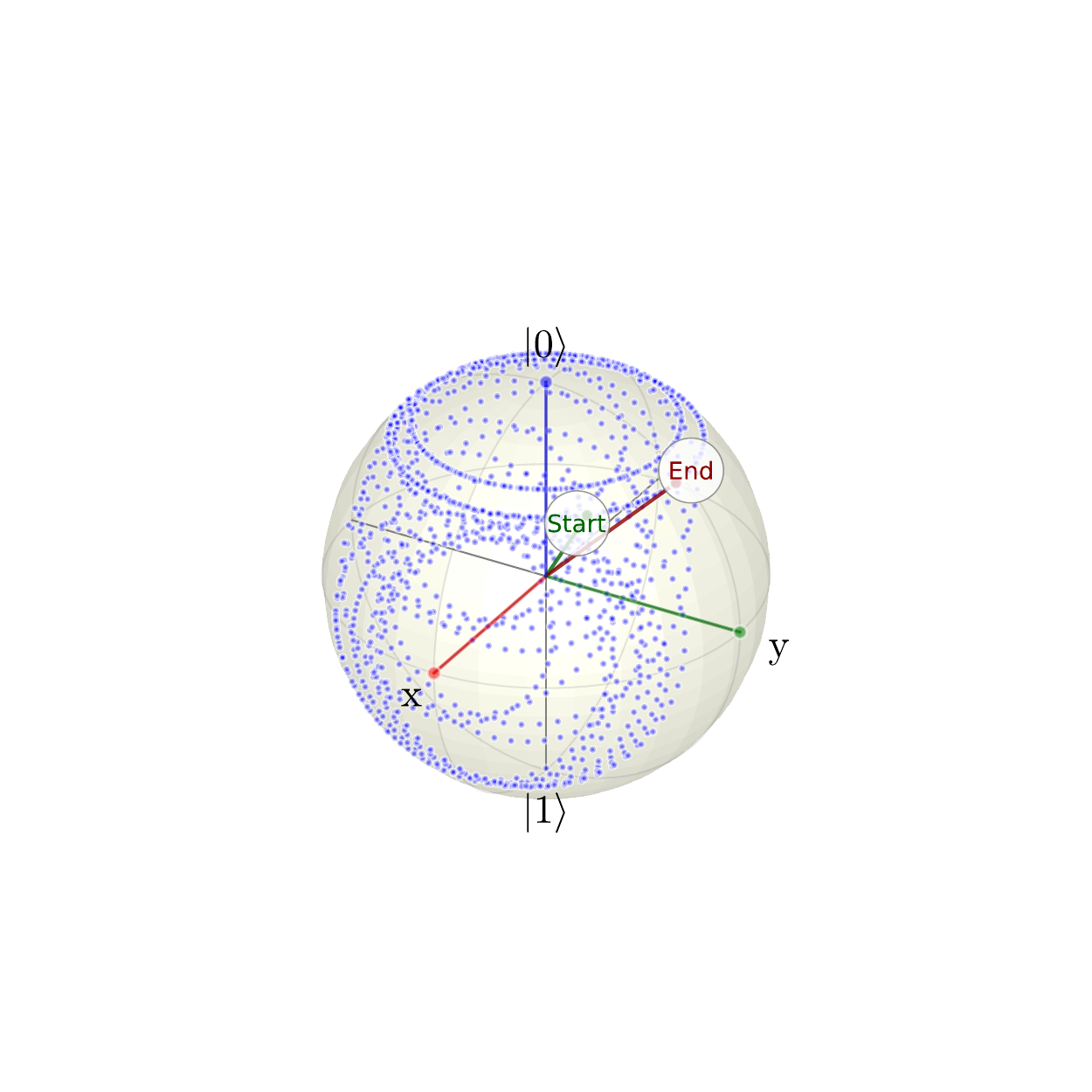}\put(0,72){\textbf{(d)}}\end{overpic}\label{fig:tgate_initial_state1_bloch}}
  \caption{T gate simulation with echo protection for different initial states. Simulation parameters and modulation functions are identical to those used in Fig.~\ref{fig:tgate}. The subfigures (a), (c) are for initial state $\theta = \pi/2$, $\phi = \pi/4$. (b), (d) are for initial state $\theta = \pi/4$, $\phi = \pi/4$. Both simulations demonstrate successful $\pi/4$ rotation around $\hat{z}$ axis, confirming the T gate's effectiveness across different initial states.}
  \label{fig:tgate_different_initial_states}
\end{figure*}

Then, as illustrated in Fig.~\ref{fig:braiding_sequence_bloch}, the resulting unitary transformation of \eq{eq:theta_braiding_sequence} is
\begin{equation}
B_{12}(\theta) = e^{\frac{\theta}{2}\gamma_1\gamma_2},
\label{eq:B_12_theta}
\end{equation}
which is the required transformation \eq{eq:theta_braiding_like_sequence} in the main text.

\section{Additional Numerical Simulation Results} \label{app:additional_simulation}
The simulation code and plotting scripts used to generate the results in the main text and appendices can be found in the repository \cite{cotunneling_braiding_plot}.

\subsection{Effects of Different Modulation Functions}

Here we take the NOT gate as an example to demonstrate the robustness of our braiding protocol across various $t_{R,\alpha \alpha'}$ modulation functions. Following Protocol~\ref{protocol:standard}, we implement the NOT gate operation given by \eq{eq:NOT_gate_B10_squared}. 
Figure~\ref{fig:different_modulations} shows the simulation results using different $t_{R,\alpha \alpha'}$ modulation functions: linear, quadratic, cubic, and cosine functions. All modulation functions successfully implement the NOT gate with high fidelity, demonstrating that our protocol is effective regardless of the specific shape of the $t_{R,\alpha \alpha'}$ modulation function. This robustness is crucial for practical implementations where the exact form of the modulation functions may vary due to experimental constraints.

\subsection{Adiabaticity and Gate Operation Time}

To investigate the adiabaticity requirements of our protocol, here following Protocol~\ref{protocol:standard}, we define $T_{\mathrm{half}}$ as the duration required to implement $B_{10}(\tfrac{\pi}{4})$, and simulate the NOT gate operation with different $T_{\mathrm{half}}$, where the system internal dynamic characteristics time is around $T_{\mathrm{dyn}}=0.3$ time units. We examine cases with $T_{\mathrm{half}} = 0.5$, $1.0$, $2.5$, and $10.0$ time units.

The results in Fig.~\ref{fig:different_T_steps} show a clear increase in gate fidelity as operation time increases, establishing that adiabaticity is well preserved when $T_{\mathrm{half}}$ is at least 10 times the system's internal dynamic characteristics time $T_{\mathrm{dyn}}$.

\subsection{T Gate Effectiveness for Different Initial States} \label{app:tgate_initial_states}

To demonstrate the effectiveness of our T gate protocol for different initial states, we performed simulations with various initial qubit states following Protocol~\ref{protocol:tgate}. Figure~\ref{fig:tgate_different_initial_states} demonstrates successful $\pi/4$ rotation around $\hat{z}$ axis for two representative initial states: $|\psi\rangle=\frac{1}{\sqrt{2}}|\uparrow\rangle+\frac{e^{i\pi/4}}{\sqrt{2}}|\downarrow\rangle$ and $|\psi\rangle=\cos(\pi/8)|\uparrow\rangle+e^{i\pi/4}\sin(\pi/8)|\downarrow\rangle$. The results confirm the T gate's effectiveness across different initial states.

\bibliography{apssamp}

\end{document}